%% file: main.tex
\documentclass[conference]{IEEEtran}

\IEEEoverridecommandlockouts                              

\usepackage{xr}

\newcommand{\omitted}[1]{}

\input{titleAndAuthors}

\usepackage{cite}

\usepackage{comment}
\usepackage{siunitx}
\usepackage{relsize}
\usepackage{ifthen}
\usepackage[colorinlistoftodos]{todonotes}

\usepackage[vlined,ruled,linesnumbered]{algorithm2e}
\usepackage{graphics} 
\usepackage{rotating}
\usepackage{color}
\usepackage{enumerate}
\usepackage[T1]{fontenc}
\usepackage{psfrag}
\usepackage{epsfig} 
\usepackage{booktabs}
\usepackage{graphicx,url}
\usepackage{multirow}
\usepackage{array}
\usepackage{latexsym}
\usepackage{amsfonts}
\usepackage{amsmath}
\usepackage{amssymb}
\usepackage{xstring}
\usepackage{algorithmic}
\usepackage{multirow}
\usepackage{xcolor}
\usepackage{prettyref}
\usepackage{flexisym}
\usepackage{bigdelim}
\usepackage{breqn} 
\usepackage{listings}
\let\labelindent\relax
\usepackage{xspace}
\usepackage{bm}
\graphicspath{{./figures/}}
\usepackage{tikz}
\usetikzlibrary{matrix,calc}
\usepackage{lipsum}
\usepackage{mdwlist}

\let\itemize\undefined
\makecompactlist{itemize}{stditemize}
\usepackage{enumitem}
\usepackage{caption}
\usepackage{amsthm}
\usepackage{mathtools}

\makeatletter
\let\NAT@parse\undefined
\makeatother
\usepackage{hyperref}
\usepackage{cleveref}

\hypersetup{%
  colorlinks=true,
  linkcolor=blue,
  filecolor=magenta,      
  urlcolor=black,
  citecolor=red,
  linkbordercolor={0 0 1}
}

\input{preamble_symbols.tex}

\input{shortcuts.tex}

\PassOptionsToPackage{end}{algorithmic}

\begin{document}

\maketitle

\thispagestyle{empty}
\pagestyle{empty}

\input{Abstract.tex}

\input{1-Introduction.tex}

\input{2-Problem.tex} 

\input{3-Algorithms.tex} 

\input{4-Guarantees.tex} 

\input{5-Experiments.tex}

\input{6-Conclusion.tex}


\bibliographystyle{IEEEtran}
\bibliography{references}

\newpage
\onecolumn 

\appendices
\input{App-1} 

\input{App-2}

\input{App-3}

\end{document}

%% file: titleAndAuthors.tex
\title{ 
Bandit Submodular Maximization for Multi-Robot Coordination 
in Unpredictable and Partially Observable Environments
}
\author{Zirui Xu$^\star$, Xiaofeng Lin$^\dagger$, Vasileios Tzoumas$^\star$ \\
$^\star$Department of Aerospace Engineering, $^{\dagger}$Department of Robotics \\
University of Michigan\\
{\tt
\{ziruixu,linxiaof,vtzoumas\}@umich.edu}
}

%% file: preamble_symbols.tex
\newtheorem{theorem}{Theorem}
\newtheorem{problem}{Problem}

\newtheorem{assumption}{Assumption}
\newtheorem{definition}{Definition}
\newtheorem{proposition}{Proposition}

\newtheorem{remark}{Remark}

\newcommand{\bdmath}{\begin{dmath}}
\newcommand{\edmath}{\end{dmath}}
\newcommand{\beq}{\begin{equation}}
\newcommand{\eeq}{\end{equation}}
\newcommand{\bdm}{\begin{displaymath}}
\newcommand{\edm}{\end{displaymath}}
\newcommand{\bea}{\begin{eqnarray}}
\newcommand{\eea}{\end{eqnarray}}
\newcommand{\beal}{\beq \begin{array}{lll}}
\newcommand{\eeal}{\end{array} \eeq}
\newcommand{\beas}{\begin{eqnarray*}}
\newcommand{\eeas}{\end{eqnarray*}}
\newcommand{\ba}{\begin{array}}
\newcommand{\ea}{\end{array}}
\newcommand{\bit}{\begin{itemize}}
\newcommand{\eit}{\end{itemize}}
\newcommand{\ben}{\begin{enumerate}}
\newcommand{\een}{\end{enumerate}}

\newcommand{\calA}{{\cal A}}
\newcommand{\calB}{{\cal B}}

\newcommand{\calN}{{\cal N}}

\newcommand{\calS}{{\cal S}}
\newcommand{\calT}{{\cal T}}

\newcommand{\calV}{{\cal V}}

\definecolor{myblue}{RGB}{65 105 225}

\newcommand{\hide}[1]{}

\newcommand{\hiddenText}{{\color{gray} hidden text.}}
\newcommand{\hideWithText}[1]{\hiddenText}

\newcommand{\opt}{^{\star}}

\newcommand{\scenario}[1]{{\fontsize{9}{9}\selectfont\sf #1}\xspace}



%% file: shortcuts.tex
\newcommand{\ie}{\emph{i.e.},\xspace}
\newcommand{\eg}{\emph{e.g.},\xspace}
\newcommand{\myin}{\, \in \,}
\newcommand{\red}[1]{{\color{red}#1}}

\newcommand{\sg}{\scenario{SG}}
\newcommand{\banalg}{\scenario{BSG}}
\newcommand{\sgh}{{\fontsize{9}{9}\selectfont\sf SG\text{-}Heuristic}}

\newcommand{\myParagraph}[1]{{\bf #1.}\xspace}

\renewcommand{\opt}{\scenario{OPT}}

\newcommand{\expsix}[1]{\scenario{EXP3$^{\star}$-SIX$|_{#1}$}}

\newcommand{\solbsg}{\calA^{\banalg}}
\newcommand{\distfsf}{p}

\newcommand{\solopt}{\calA^{\opt}}

\newcommand{\obsenv}{E_t^{obs}}
\newcommand{\env}{E_t}

%% file: Abstract.tex
\begin{abstract}%
We study the problem of multi-agent coordination in \textit{unpredictable \emph{and} partially observable} environments, that is, environments whose future evolution is unknown~a priori and that can only be partially observed. We are motivated by the future of autonomy that involves multiple robots coordinating actions in dynamic, unstructured, and partially observable environments to complete  complex tasks such as target tracking, {environmental mapping}, and area monitoring.  Such tasks are often modeled as submodular maximization coordination problems due to the information overlap among the robots. We introduce the first submodular coordination algorithm with {bandit feedback} and bounded {tracking regret} ---\textit{bandit feedback} is the robots' ability to compute in hindsight only the effect of their chosen actions, instead of all the alternative actions that they could have chosen instead, due to the partial observability; and \textit{tracking regret} is the algorithm's suboptimality with respect to the optimal time-varying actions that fully know the future a priori. The bound gracefully degrades with the environments' capacity to change adversarially, quantifying how often the robots should re-select actions to learn to coordinate as if they fully knew the future a priori. The algorithm generalizes the seminal {Sequential Greedy} algorithm  by Fisher et al.~to the bandit setting, by leveraging submodularity and algorithms for the problem of \textit{tracking the best action}. We validate our algorithm in simulated scenarios of multi-target tracking.
\end{abstract}

%% file: 1-Introduction.tex
\section{Introduction}\label{sec:Intro}

In the future, autonomous robots will be collaboratively planning actions in complex tasks such as {target tracking}~\cite{corah2021scalable}, {environmental mapping}~\cite{atanasov2015decentralized}, and {area monitoring}~\cite{xu2022resource}. Such multi-robot tasks have been modeled by researchers in robotics and control via maximization problems of the form
\begin{equation}\label{eq:intro}
	\max_{a_{i,\,t}\,\in\,\mathcal{V}_i,\,  \forall\, i\,\in\, \calN}\
	f_t(\,\{a_{i,\,t}\}_{i\myin \calN}\,), \ \ t=1,2,\ldots,
\end{equation}
where $\calN$ is the robot set, $a_{i,\,t}$ is robot $i$'s action at time step $t$, $\calV_i$ is robot $i$'s set of available actions, and $f_t:2^{\prod_{i \in \calN}\calV_i}\mapsto\mathbb{R}$ is the objective function that captures the task utility at time step $t$.  Specifically, the objective function $f_t$ is considered computable prior to each time step~$t$ given a model about the future evolution of the environment~\cite{krause2008near,singh2009efficient,tokekar2014multi,atanasov2015decentralized,gharesifard2017distributed,grimsman2018impact,corah2018distributed,corah2019distributed,corah2021scalable,schlotfeldt2021resilient,xu2022resource}; 
\eg in target tracking, a stochastic model for the targets' future motion is often considered available, and then $f_t$ can be chosen for example as the mutual information between the position of the robots and that of the targets~\cite{atanasov2015decentralized}.

The optimization problem in~\cref{eq:intro} is NP-hard~\cite{Feige:1998:TLN:285055.285059} but near-optimal approximation algorithms are possible in polynomial time when $f_t$ has a special structure, especially, when $f_t$ is \textit{submodular}~\cite{fisher1978analysis} ---submodularity is a diminishing returns property, and in multi-robot information gathering tasks it emanates due to the possible information overlap among the information gathered by the robots~\cite{krause2012submodular}. One celebrated  approximation algorithm for~\cref{eq:intro} is the \textit{Sequential Greedy} algorithm~\cite{fisher1978analysis}, which achieves a near-optimal $1/2$ approximation bound when $f_t$ is submodular.  All the above multi-robot tasks and more, from target tracking and environmental exploration to collaborative mapping and area monitoring, can be modeled as submodular coordination problems, and thus, Sequential Greedy and its variants have been commonly used in robotics~\cite{krause2008near,singh2009efficient,tokekar2014multi,atanasov2015decentralized,gharesifard2017distributed,grimsman2018impact,corah2018distributed,corah2019distributed,corah2021scalable,schlotfeldt2021resilient,liu2021distributed,robey2021optimal,rezazadeh2023distributed,konda2022execution,krause2012submodular,xu2022resource}. 

But the application of the Sequential Greedy algorithm and its variants can be hindered in challenging environments that are unpredictable and partially observable:
\begin{figure}[t]
    \captionsetup{font=footnotesize}
    \centering
    \includegraphics[width=.98\columnwidth]{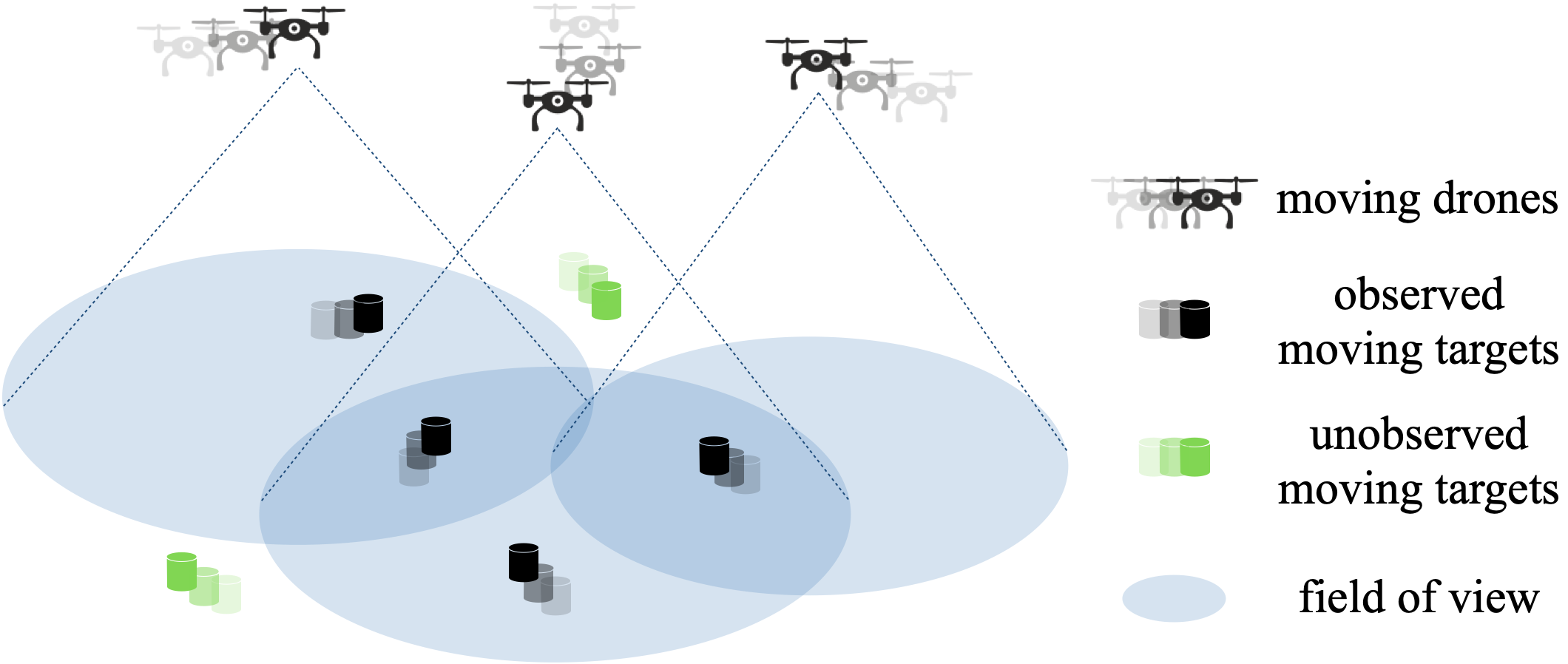}\\
    \caption{\textbf{Example of Multi-Robot Coordination in Unpredictable and Partially Observable Environments: Target Tracking.} In this paper, we focus on multi-robot coordination tasks where the robots' capacity to select effective actions  is compromised by (i) a lack of knowledge about how the environment will evolve, and (ii) a lack of full observability of the environment's evolution.  For example, in target tracking, drones are often tasked to coordinate their motion to maximize at each time step the number of tracked targets.  But in adversarial scenarios, (i) the robots may be unaware of the targets' intentions and motion model, thus being unable to plan effective actions by simulating the future, and (ii) the robots may carry sensors with a limited field of view, thus being unable to reason even in hindsight whether alternative actions could have been more effective in tracking targets.
   Notwithstanding the said challenges, in this paper, we aim to provide a general-purpose coordination algorithm achieving bounded suboptimality against the optimal multi-robot actions in hindsight.}\label{fig:partial}
   \vspace{-7mm}
\end{figure}

\paragraph{Unpredictable Environments} The said complex tasks often evolve in environments that change unpredictably, \ie in environments whose future evolution is unknown a priori.  For example, during target tracking the targets' actions can be unpredictable when their intentions and maneuvering capacity are unknown~\cite{sun2020gaussian}. 
In such challenging environments, the robots that are tasked to track the targets cannot simulate the future to compute $f_t$ prior to time step $t$, \ie the robots cannot utilize the Sequential Greedy algorithms and its variants~\cite{krause2008near,singh2009efficient,tokekar2014multi,atanasov2015decentralized,gharesifard2017distributed,grimsman2018impact,corah2018distributed,corah2019distributed,corah2021scalable,schlotfeldt2021resilient,xu2022resource}. Instead, the robots have to coordinate their actions online by relying on past information only, \ie by relying only on the retrospective reward of their actions upon the observation of the environment's evolution.

Such online coordination algorithms have been recently proposed in~\cite{xu2023online}. Particularly, \cite{xu2023online} provides a submodular coordination algorithm with guaranteed suboptimality against the robots' optimal \textit{time-varying} actions in hindsight ---the optimal actions ought to be time-varying to be effective against a changing environment such as an evading target.\footnote{Additional algorithms have been proposed for the case where in \cref{eq:intro} $f_t$ is unknown a priori but these algorithms apply to tasks where the optimal actions are \textit{static}~\cite{streeter2008online,streeter2009online,suehiro2012online,golovin2014online,chen2018online,zhang2019online}, instead of time-varying, guaranteeing bounded suboptimality with respect to optimal time-\underline{in}variant actions.}

\paragraph{Partially Observable Environments} But online coordination methods such as~\cite{xu2023online} become {inapplicable} when the unpredictable environments are {only} partially observable: when an environment is partially observable, online learning methods can only compute the utility of their {executed} actions. Particularly, they cannot compute in hindsight the utility of actions they could have selected alternatively to execute ---this partial-information feedback is known as \textit{bandit feedback}~\cite{lattimore2020bandit}.  Bandit feedback hence compromises the capacity of online learning methods to learn near-optimal action policies based on past information only.
Take the target tracking scenario in Fig.~\ref{fig:partial} as an example: since the drones have limited fields of view, they can observe only part of the targets (those inside the field of view), being unaware of any unobserved targets (those outside the field of view); consequently, the robots cannot compute even in hindsight how many targets they would have seen instead if they had chosen alternative actions. 

Although recent contributions~\cite{baykal2017persistent,zhang2019partially,lee2021upper,landgren2021distributed,dahiya2022scalable,wakayama2022active} have focused on multi-robot coordination subject to bandit feedback, they consider partially observable environments where (i) the environments' state evolution is governed by an unknown stochastic i.i.d model, 
and (ii) the robots' goal is to learn actions that maximize the sum of the robots' individual rewards without accounting for the possible information overlap among the information gathered by the robots; \eg in the context of Fig.~\ref{fig:partial}, the sum of the robots' individual rewards is $6$ since the drones on the left, center, and right observe $2$, $3$, and $1$ targets, respectively.

\myParagraph{Goal} In this paper, we focus instead on unpredictable and partially observable environments where: (i) the environments' state is non-stochastic and even adversarial, {that is, the environment’s behavior is not governed by a probability model and can even be adaptive to the robots’ action,} \eg where the robots are tasked to track targets that can adapt their motion to the robots' motion; and~~(ii) the robots' goal is to learn actions for each time step $t$ that maximize a global objective $f_t$ that is submodular, instead of a mere addition of the individual rewards of the robots.  Accounting for the submodularity structure is crucial since it quantifies the possible information overlap among the information gathered by the robots; \eg in the context of Fig.~\ref{fig:partial}, the number of tracked targets by the drones is $4$, instead of $6$ as we computed above when we ignored the information overlap.

\myParagraph{Contributions}
We provide the first bandit submodular optimization algorithm for multi-robot coordination in unpredictable and partially observable environments (\Cref{sec:algorithm}).  We name the algorithm \textit{Bandit Sequential Greedy} (\banalg).  
\banalg generalizes the Sequential Greedy algorithm~\cite{fisher1978analysis} from the setting where each $f_t$ is fully known a priori to the bandit setting. 
\banalg has the properties:
\begin{itemize}[leftmargin=9pt]
	\item \textit{Computational Complexity}: {For each agent $i$, \banalg requires only one function evaluation and $O(\log{T})$ additions and multiplications per agent per round } (\Cref{subsec:complexity}).
	
	\item \textit{Approximation Performance}: \banalg guarantees bounded \textit{tracking regret} (\Cref{subsec:tracking-regret}), \ie bounded suboptimality with respect to optimal time-varying actions that know the future a priori.  The bound gracefully degrades with the environments' capacity to change adversarially, quantifying how often the robots should re-select actions to learn to coordinate as if they knew the future a priori. In more detail, the bound guarantees that the agents select actions that asymptotically and in expectation match the near-optimal performance of the Sequential Greedy algorithm~\cite{fisher1978analysis} in known environments. 
\end{itemize}

To enable \banalg, we make the technical contributions:

\quad \textit{1)}~~First, we enable each robot to retrospectively estimate the reward of all its available actions despite the bandit feedback.  To this end, we 
use as a subroutine on-board each robot a novel algorithm for the problem \textit{tracking the best action with bandit feedback}~\cite{auer2002nonstochastic} ---we are inspired to this end by~\cite{xu2023online,matsuoka2021tracking}, which leverage similar subroutines for online submodular optimization problems in fully observable environments.\footnote{\hspace{-.001mm}\cite{xu2023online} focuses on online submodular coordination in fully observable environments, instead of partially observable environments. Further, \cite{matsuoka2021tracking} focuses on the problem of \textit{cardinality-constrained submodular maximization in fully observable environments}, which has the form $\max_{\calS\,\subseteq\, \calV, \, |\calS|\,\leq\, k}\, f(\calS)$, given an integer $k$ and an $f:2^\calV\mapsto \mathbb{R}$, and is thus distinct from \cref{eq:intro}.} 
Particularly, although the current algorithm for the problem of \textit{tracking the best action with bandit feedback}, namely, \scenario{EXP3-SIX} \cite{neu2015explore}, can guarantee a bounded tracking regret for that problem, it requires the a priori knowledge of a parameter capturing how fast the environment is going to change.  Satisfying such a requirement is typically infeasible in practice.   Therefore, in this paper, we use a ``doubling trick''~\cite{shalev2012online} to extend \scenario{EXP3-SIX} to an algorithm that requires no more the a priori knowledge of this parameter (\Cref{subsec:tracking-best-sequence}); we name the algorithm \scenario{EXP3$^\star$-SIX} (\Cref{alg:EXP3star-SIX}).  

\quad \textit{2)}~~Then, we leverage (i) \scenario{EXP3$^\star$-SIX}'s regret guarantee for the problem of tracking the best action with bandit feedback (\Cref{th:EXP3star-SIX}), (ii) \banalg's steps 
(\Cref{alg:online}), and (iii) $f_t$'s submodularity to prove \banalg's tracking regret guarantee for the coordination problem of this paper (Appendix). 

\myParagraph{Numerical Evaluations} 
We evaluate \banalg in simulated scenarios of target tracking with multiple robots (\Cref{sec:experiments}), where the robots carry noisy sensors with limited field of view to observe the targets.  We consider scenarios where $2$ robots pursue $2$, $3$, or $4$ targets.  For each scenario, we first consider non-adversarial targets and, then, adversarial targets: the non-adversarial targets traverse predefined trajectories, independently of the robots' motion; whereas, the adversarial targets maneuver in response to the robots' motion.  \textit{In both cases, the targets' future motion and maneuvering capacity are unknown to the robots.} 
Across the simulations, \banalg encourages the robots to maximize their tracking capability,
also enabling collaborative behaviors such as robots switching targets to improve speed compatibility (fast robot vs.~fast target, {instead of slow robot vs.~fast target).} 

%% file: 2-Problem.tex
\section{Bandit Submodular Coordination with Bounded Tracking-Regret}\label{sec:problem}

We define the problem \textit{Bandit Submodular Coordination}. To this end, we use the notation:
\begin{itemize}
    \item $\calV_\calN \triangleq \prod_{i\myin \calN} \,\calV_i$ is the cross product of sets $\{\calV_i\}_{i\myin \calN}$.
    \item $[T]\triangleq\{1,\dots,T\}$ for any positive integer $T$;
    \item $f(\,a\,|\,\calA\,)\triangleq f(\,\calA \cup \{a\}\,)-f(\,\calA\,)$ is the marginal gain of set function $f:2^\calV\mapsto \mathbb{R}$ for adding $a \in \calV$ to $\calA \subseteq\calV$.
    \item $|\calA|$ is the cardinality of a discrete set $\calA$. 
\end{itemize}

The following preliminary framework is also considered.

\myParagraph{Agents}  $\calN$ is the set of all agents ---the terms ``\emph{agent}'' and ``\emph{robot}'' are used interchangeably in this paper.  The agents coordinate actions to complete a task.  To this end, they can observe one another’s selected actions at each time step. 

\myParagraph{Actions} $\calV_i$ is a \textit{discrete} and \textit{finite} set of actions available and always known to robot $i$.  For example, $\calV_i$ may be a set of motion primitives  that robot $i$ can execute to move in the environment~\cite{tokekar2014multi} or robot $i$'s discretized control inputs~\cite{atanasov2015decentralized}. 

\myParagraph{Environment} $E_t$ is the state of the  environment at time step $t$. $E_t$  {evolves} (possibly adversarially) with the agents' past actions up to $t-1$.  Also, $\env$ is unpredictable prior to time $t$, in particular, the robots are unaware of a model capturing the future evolution of the environment. For example, in the multi-target tracking  scenario in Fig.~\ref{fig:partial}, where the robots have \underline{no} model about the future motion of the targets, at time $t-1$ the robots cannot know where the targets will be at time $t$.

\myParagraph{Observable Environment}
$\obsenv(\{a_{i,\,t}\}_{i\myin \calN})$ is the part of $\env$ observed by the robots once the robots have executed their actions $\{a_{i,\,t}\}_{i\myin \calN}$ at time step $t$. For example, in Fig.~\ref{fig:partial}, while $\env$ includes all targets' positions, $\obsenv(\{a_{i,\,t}\}_{i\myin \calN})$ includes only the positions of the targets that are within the collaborative field of view (black-colored targets).

\myParagraph{Objective Function} 
The agents coordinate their actions $\{a_{i,\,t}\}_{i\myin \calN}$ to maximize an objective function $f(\{a_{i,\,t}\}_{i\myin \calN},\, \env)$ ---{we explicitly note the dependence of the value $f$ of the actions on the state of the environment}. In Fig.~\ref{fig:partial} for example, $f$ is equal to $4$ since four targets are within the field of view of the robots given the robots' and targets' positions at time $t$. We henceforth consider
\begin{equation}\label{eq:f_t}
	f_t(\cdot)\triangleq f(\cdot,\, \env).
\end{equation}

\myParagraph{Bandit Feedback} $f(\cdot,\, \env)$ is unknown prior to the execution of the actions $\{a_{i,\,t}\}_{i\myin \calN}$ since $E_t$ is unknown before time $t$. Upon the execution of the actions $\{a_{i,\,t}\}_{i\myin \calN}$, if $\env$ is fully observable, then the robots can evaluate $f(\calA,\, \env)$ for all $\calA\subseteq\{\calV_i\}_{i\myin \calN}$; \ie the robots can evaluate in hindsight the performance of any actions that they could have chosen instead for time $t$.  But in this paper, the environment is generally partially observable, hence,  upon the execution of the actions $\{a_{i,\,t}\}_{i\myin \calN}$, the robots can only evaluate $f(\calA,\, \obsenv(\calA))$ for all $\calA\subseteq\{a_{i,\,t}\}_{i\myin \calN}$. We refer to the two said cases of information feedback as \textit{full feedback} and \textit{bandit feedback}, per similar definitions in the literature of online learning and optimization~\cite{lattimore2020bandit}.

\begin{assumption}[Exact Function Evaluation Despite Partially Observable Environments]\label{ass:partial}
We assume coordination tasks where for all $\calA\subseteq\{\calV_i\}_{i\myin \calN}$,
    \begin{equation}\label{eq:partial}
        f(\calA,\, \obsenv(\calA))\equiv f(\calA,\, E_t).
    \end{equation}
\end{assumption}

Coordination tasks that satisfy \Cref{ass:partial} include target tracking \cite{corah2021scalable}, environmental mapping \cite{atanasov2015decentralized}, and area monitoring \cite{schlotfeldt2021resilient}, where, intuitively, the objective function is defined based on observed information only; \eg in the multi-target tracking scenario in Fig.~\ref{fig:partial}, the robots know exactly how many targets are within their field of view, thus \Cref{ass:partial} holds true when $f$ is the number of targets within the robots' field of view.  In contrast, \Cref{ass:partial} would be violated if the objective in Fig.~\ref{fig:partial} were to minimize the distance between each target and its nearest drone: then, the drones cannot possibly evaluate their distance to the unobservable targets (colored green in the figure) and, hence, \Cref{ass:partial} cannot hold true. {In all, satisfying \Cref{ass:partial} implies that what the robots perceive onboard exactly equals what they really achieve in a partially observable environment. }

\myParagraph{Submodular Structure} In information gathering tasks
such as target tracking, environmental mapping, and area monitoring, typical objective functions are the covering functions~\cite{corah2021scalable,atanasov2015decentralized,xu2022resource}. These functions capture how much
area/information is covered given the actions of all robots.
They satisfy the properties defined below (Definition~\ref{def:submodular}).

\begin{definition}[Normalized and Non-Decreasing Submodular Set Function{~\cite{fisher1978analysis}}]\label{def:submodular}
A set function $f:2^\calV\mapsto \mathbb{R}$ is \emph{normalized and non-decreasing submodular} if and only if 
\begin{itemize}
\item (Normalization) $f(\,\emptyset\,)=0$;
\item (Monotonicity) $f(\,\calA\,)\leq f(\,\calB\,)$, $\forall\,\calA\subseteq \calB\subseteq \calV$;
\item (Submodularity) $f(\,s\,|\,\calA\,)\geq f(\,s\,|\,{\mathcal{B}}\,)$, $\forall\,\calA\subseteq {\mathcal{B}}\subseteq\calV$ and $s\in \calV$.
\end{itemize}
\end{definition}

Normalization holds without loss of generality.  In contrast, monotonicity and submodularity are intrinsic to the function.  
Intuitively, if $f(\,\calA\,)$ captures the number of targets tracked by a set $\calA$ of drones, then the more drones are deployed, no fewer targets are covered; this is the monotonically non-decreasing property.  Also, the marginal gain of tracked targets caused by deploying a drone $s$ \emph{drops} when \emph{more} drones are already deployed; this is the submodularity~property.

\myParagraph{Problem Definition}  In this paper, we focus on:
\begin{problem}[Bandit Submodular Coordination]
\label{pr:online}
Assume a time horizon $H$ of operation discretized to $T$ time steps. At each time step $t\in[T]$, the agents $\calN$ select actions $\{a_{i,\;t}\}_{i\,\in\,\calN}$ \emph{online} such that  they solve 
\begin{equation}
    \max_{a_{i,\,t}\,\in\,\mathcal{V}_i, \,
         \forall\, i\,\in\, \calN}\
    f_t(\,\{a_{i,\,t}\}_{i\myin \calN}\,),
\end{equation}
where $f_t: 2^{\prod_{i\myin \calN} \,\calV_i}\mapsto \mathbb{R}$ is a normalized and non-decreasing submodular set function, and the agents can access the values of $f_t(\,\calA\,)$ only after they have selected $\{a_{i,\,t}\}_{i\myin \calN}$, $\forall\,\calA\subseteq\{a_{i,\,t}\}_{i\myin \calN}$.
\end{problem}

\begin{remark}[Adversarial Environment and Randomized Algorithm]\label{rem:adversarial}
Dependent on the agents' past actions, the environment $\env$ in \Cref{pr:online} can be \emph{adversarial}, in that it can decide $f_t$ at each time step $t$ to change for worsening the reward of $\{a_{i,\;t}\}_{i\,\in\,\calN}$. $E_t$ here serves as the {\emph{adversary}} in a bandit problem \cite{slivkins2019introduction}. In this paper, we provide a randomized algorithm that guarantees in expectation a suboptimality bound that degenerates gracefully as the environment becomes more adversarial. If $E_t$ makes $f_t$ change arbitrarily much between consecutive time steps, then inevitably no algorithm can guarantee a near-optimal performance.
\end{remark}

%% file: 3-Algorithms.tex
\section{Bandit Sequential Greedy ({\fontsize{9}{9}\selectfont\sf BSG}) Algorithm} \label{sec:algorithm}

We present the Bandit Sequential Greedy (\banalg) algorithm for \Cref{pr:online}. \banalg leverages
as subroutine an algorithm we introduce for the problem of \textit{tracking the best action with bandit feedback}.  Thus, before we present \banalg in \Cref{subsec:bsg}, we first present the algorithm for \textit{tracking the best action with bandit feedback} in \Cref{subsec:tracking-best-sequence}.

\subsection{The \scenario{EXP3$^\star$-SIX} Algorithm for Tracking the Best Action with Bandit Feedback}\label{subsec:tracking-best-sequence}

\input{alg-EXP3star-SIX}

\textit{Tracking the best action with bandit feedback} is an \textit{adversarial bandit} problem \cite{lattimore2020bandit}. It involves an agent {---instead of the entire team---} selecting a sequence of actions to maximize the total reward over a given number of time steps.  The challenge is dual: (i) the reward associated with each action is decided by the environment at each time step and unknown to the agent until the action has been executed; and (ii) the agent receives only  bandit feedback of the rewards. To solve the problem, the agent needs to leverage past observation of the rewards till the last time step to predict the best action that achieves the highest reward for this time step.

To formally state the problem, we use the notation:
\begin{itemize}
    \item $\calV$ denotes the agents' available action set;
    \item $a_{t}\in\calV$ denotes the agent's selected action at time $t$;
    \item $a_t^\star$ denotes the best action that achieves the highest reward among $\calV$ at $t$;
    \item $r_{a_t,\,t}\in[0,1]$ denotes the reward that the agent receives by selecting action $a_{t}$ at $t$;
    \item $\tilde{r}_{t}\in[0,1]^{|\calV|}$ denotes the estimation of the rewards of all actions available to the agent at $t$;
    \item ${\bf 1}(\cdot)$ is the indicator function, \ie ${\bf 1}(x)=1$ if the event $x$ is true, otherwise ${\bf 1}(x)=0$.
    \item $P(T) \triangleq \sum_{t=1}^{T-1} {\bf 1}(a^\star_t\neq a^\star_{t+1})$ counts how many times the best action changes over $T$ time steps due to the {adversary (the environment)}.
\end{itemize}

\begin{problem}[Tracking the Best Action with Bandit Feedback \cite{neu2015explore}]\label{pr:tracking-the-best-action}
Assume a time horizon $H$ of operation discretized to $T$ time steps. The agent selects an action $a_{t}$ \emph{online} at each time step $t\in[T]$ to solve the optimization problem
\begin{equation}
    \max_{a_{t}\myin\mathcal{V},\, t\myin[T]} \;\; \sum_{t=1}^T\;r_{a_t,\,t},
\end{equation}
where only the reward $r_{a_t,\,t}\in[0,1]$ becomes known to the agent and only \emph{once} $a_{t}$ has been selected.
\end{problem}

A randomized algorithm is needed for \Cref{pr:tracking-the-best-action}, given that the environment can adversarially adapt to the agent's previously selected actions to seek to minimize the agent's total reward. If an algorithm for \Cref{pr:tracking-the-best-action} is deterministic, then the environment can know a priori the action $a_t$ to be selected by the deterministic algorithm for each time step $t$ and accordingly choose the rewards $r_{a_t,\,t}=0$ and $r_{a_t',\,t}=1$, $\forall a_t'\in\calV, a_t'\neq a_t$. This will lead to $\sum_{t=1}^{T} r_{a_t^\star,\,t} - r_{a_t,\,t} \geq T(1-1/|\calV|)$, which means $a_t$ can never converge to $a_t^\star$, $\forall t\in[T]$ \cite[Chapter 11.1]{lattimore2020bandit}. Therefore, at each time step $t$, we need a randomized algorithm to provide a probability distribution $p_t$ over the action set $\calV$, from which the agent can draw the action $a_t$ for time step $t$.

Moreover, a desired randomized algorithm for \Cref{pr:tracking-the-best-action} should  ensure $\mathbb{E}\left[\sum_{t=1}^{T} r_{a_t^\star,\,t} - r_{a_t,\,t}\right] = \sum_{t=1}^{T} r_{a^\star_t,\,t} - r_{t}^{\top}p_t$ is sublinear, where the expectation results from the internal randomness of the algorithm, such that as $T\to\infty$, $\frac{1}{T}\mathbb{E}\left[\sum_{t=1}^{T} r_{a_t^\star,\,t} - r_{a_t,\,t}\right] \to 0$, and thus $a_t\to a_t^\star$.

Although \scenario{EXP3-SIX} \cite{neu2015explore} is an algorithm that achieves a sublinear $\sum_{t=1}^{T} r_{a^\star_t,\,t} - r_{t}^{\top}p_t$, it requires the value of $P(T)$ {for picking a ``learning rate'' that can bound the suboptimality. The learning rate decides how fast the algorithm adapts to the environmental change}. But $P(T)$ is unknown a priori.  Thus, in this paper we leverage a ``doubling trick'' technique, common in the literature of online learning \cite{zhang2018adaptive,matsuoka2021tracking}, 
and present a new algorithm, \scenario{EXP3$^\star$-SIX}, that overcomes \scenario{EXP3-SIX}'s said limitation. Specifically, \scenario{EXP3$^\star$-SIX} {uses the multiplicative weight update (MWU) method \cite{cesa2006prediction} to synthesize the results of multiple subroutines of \scenario{EXP3-SIX} with different learning rates (lines~9-16), at least one of which is close enough to the learning rate computed using $P(T)$}. 

In more detail, \Cref{alg:EXP3star-SIX} initializes and maintains $J$ subroutines of \scenario{EXP3-SIX}, each associated with a weight $z_{j,\,t}$ and a different learning rate $\eta^{(j)}$, $j\in[J]$. For each $j\in[J]$,     a weight $w_{i,\,t}^{(j)}$ is assigned to each available action $i\in\calV$ (lines 1-4). 
At each time step $t\in[T]$,  \Cref{alg:EXP3star-SIX} first uses MWU to compute the probability distribution $p_t$ based on $\{w_{i,\,t}^{(j)}\}_{j\myin[J]}$ and $\{z_{j,\,t}\}_{j\myin[J]}$ (lines 5-6). 
Then, after outputting $p_t$ and observing the new reward $r_{a_t,\,t}$  (lines 7-8), \Cref{alg:EXP3star-SIX} computes an estimate $\{\tilde{r}_t^{(j)}\}_{j\myin[J]}$ for all available actions' rewards (lines 9-11). $\{\tilde{r}_t^{(j)}\}_{j\myin[J]}$ is an optimistically biased estimate of $r_{a_t,\,t}$, \ie larger than the unbiased estimate. {The smaller is $r_{a_t,\,t}$, the larger is $(\gamma^{(j)}\,(1 - r_{a_t,\,t})) / (p_{i,\,t}\,(p_{i,\,t} + \gamma^{(j)}))$, \ie the amount of the bias of $\tilde{r}_t^{(j)}$. Therefore, the smaller is $r_{a_t,\,t}$, the more \Cref{alg:EXP3star-SIX} is encouraged to explore.} {Finally, for the next time step, \Cref{alg:EXP3star-SIX} updates $\{w_{i,\,t+1}^{(j)}\}_{j\myin[J]}$ with the Fixed Share Forecaster~\cite{cesa2006prediction} steps (lines 12-14) and obtains $\{z_{j,\,t+1}\}_{j\myin[J]}$ using MWU (line 15).}
The higher is $\eta^{(j)}$, the more $\{w_{i,\,t}^{(j)}\}_{j\in[J]}$ depend on the recently observed rewards and, thus, the faster \scenario{EXP3-SIX} $j$ adapts to the environment changes.

\begin{theorem}[Performance Guarantee of \scenario{EXP3$^\star$-SIX}]\label{th:EXP3star-SIX}
For \Cref{pr:tracking-the-best-action}, \scenario{EXP3$^\star$-SIX} guarantees
{\small
\begin{align}\label{aux:tracking_sec}
    &\sum_{t=1}^{T} r_{a^\star_t,\,t} - r_{t}^{\top}p_t  \\ \nonumber
    &\leq \Tilde{O}\left[\sqrt{T|\calV|({P}(T)+1)}\right] + \log{\left(\frac{1}{\delta}\right)\left[\Tilde{O}\left(\sqrt{\frac{T|\calV|}{{P}(T)+1}} \right)+1\right]}
\end{align}}with probability at least $1-\delta$, where $\delta\in(0,1)$ is the confidence level, and $\tilde{O}(\cdot)$ hides $\log$~terms.
\end{theorem}
\Cref{th:EXP3star-SIX} implies  $\frac{1}{T}\sum_{t=1}^{T} r_{a^\star_t,\,t} - r_{t}^{\top}p_t\to 0$ as $T\to \infty$ when $P(T)$ is sublinear in $T$, that is $P(T)/T\rightarrow 0$ for $T\rightarrow +\infty$, \ie when the optimal action $a^\star_t$ does not change too frequently across time steps; \ie in expectation the agent is able to track the best sequence of actions with high probability as $T$ increases.\footnote{\scenario{\smaller EXP3$^\star$-SIX}'s suboptimality bound in \Cref{th:EXP3star-SIX} is of the same $\tilde{O}(\cdot)$-order as  \scenario{\smaller EXP3-SIX}'s bound, despite \scenario{\smaller EXP3$^\star$-SIX} not knowing $P(T)$ a priori: in \Cref{th:EXP3star-SIX}'s proof, we show \scenario{\smaller EXP3$^\star$-SIX}'s bound contains only additional $\log$ terms with respect to $T$ and $|\calV|$ when compared to \scenario{\smaller EXP3-SIX}'s bound.}$^,$\footnote{The term $\log\left(\frac{1}{\delta}\right)\left[\Tilde{O}\left(\sqrt{T|\calV|/{(P(T)+1)}} \right)+1\right]$ in \cref{aux:tracking_sec} is always sublinear in $T$ since it is bounded by $\log\left(\frac{1}{\delta}\right)\left[\tilde{O}\left(\sqrt{T|\calV|} \right)+1\right]$.}

\subsection{The \banalg Algorithm}\label{subsec:bsg}

We present \banalg (\Cref{alg:online}). 
\banalg generalizes the Sequential Greedy 
(\sg) algorithm~\cite{fisher1978analysis} to the online setting of \Cref{pr:online}, leveraging at the agent level \scenario{EXP3$^\star$-SIX}. Particularly, when $f_t$ is known a priori, instead of unknown per \Cref{pr:online}, then \sg instructs the agents to sequentially select actions $\{a_{i,\,t}^\sg\}_{i\,\in\,\calN}$ at each $t-1$ such that
\begin{equation}\label{eq:sga}
  a_{i,\,t}^\sg \,\in\, \max_{a\myin\calV_i}\;\; f_t(\,a\;|\;\{a_{1,\,t}^\sg,\ldots,a_{i-1,\,t}^\sg\}\,),
\end{equation}
\ie agent $i$ selects $a_{i,\,t}^\sg$ after agent $i-1$, given the actions of all previous agents  $\{1,\ldots, i-1\}$, such that $a_{i,\,t}^\sg$ maximizes the marginal reward given the actions of all previous agents from $1$ to $i-1$. But since $f_t$ is unknown and even adversarial per \Cref{pr:online}, \banalg  replaces the deterministic action-selection rule of \cref{eq:sga} with a \textit{tracking the best action} rule (cf.~\Cref{rem:adversarial}). Thus, \banalg is also a sequential algorithm.

\input{alg-BSG}

\banalg starts by instructing each agent $i\in \calN$ to initialize an \scenario{EXP3$^\star$-SIX} ---we denote the \scenario{EXP3$^\star$-SIX} onboard for each agent $i$ as \expsix{i}.  Agent $i$ initializes \expsix{i} with the number $T$ of total time steps and with its action set $\calV_i$ as inputs (line 1).  Then, at each time step $t\in[T]$, in sequence:
\begin{itemize}
    \item Each agent $i$ draws an action $a_{i,\,t}^\banalg$ given the  probability distribution $\distfsf_t^{(i)}$ output by \expsix{i} (lines 5-8).
    \item All agents execute their actions $\{a_{i,\, t}^\banalg\}_{i\,\in\,\calN}$ (line 9).
    \item Each agent $i$ receives from agent $i-1$ the actions of all agents with a lower index,  $\solbsg_{i-1,\,t}$,  and then observes $f_t(\,\solbsg_{i,\,t}\,)$ (lines 10-13).
    \item  Finally, each agent $i$ computes $r_{i,\, t}^\banalg$, the reward (marginal gain) of $a_{i,\,t}^\banalg$ given $\solbsg_{i-1,\,t}$, and inputs $r_{i,\, t}^\banalg$ to \expsix{i} per line 8 of \Cref{alg:EXP3star-SIX} (lines 14-15).  With this input, \expsix{i} will compute $\distfsf_{t+1}^{(i)}$, \ie the probability distribution over the agent $i$'s actions for time step $t+1$.
\end{itemize}

%% file: alg-EXP3star-SIX.tex
\setlength{\textfloatsep}{3mm}
\begin{algorithm}[t]
	\caption{\scenario{EXP3$^\star$-SIX}.
	}\label{alg:EXP3star-SIX}
	\begin{algorithmic}[1]
		\REQUIRE Number of time steps $T$ and action set $\calV$.
		\ENSURE Probability distribution $p_t \myin \{[0,1]^{|\calV|}:\|p_t \|_1=$ $1\}$ over the actions in $\calV$ at each $t\in[T]$.
		\smallskip
		
		\STATE $J\gets\lceil\log_2{T}\rceil$, $\eta\gets\sqrt{{\log{J}}\,/\,{(2T)}}$, $\beta\gets{1}\,/\,{(T-1)}$;
		\STATE  $\eta^{(j)}\gets\sqrt{{\log{(|\calV|\,T)}}\,/\,{(2^{j-1}\,|\calV|)}}$, $\gamma^{(j)}=\eta^{(j)}/2$, for all $j\in[J]$;
		\STATE $z_{1}\gets\left[z_{1,\,1},\dots,z_{J,\,1}\right]^\top$ with $z_{j,\,1}=1$, for all $j\in[J]$;
		\STATE $w_{1}^{(j)}\gets\left[w_{1,\,1}^{(j)},\dots,w_{|\calV|,\,1}^{(j)}\right]^\top$ with $w_{i,\,1}^{(j)}=1$, for all  $t\in[T]$ and $i\in\calV$;
        \FOR {each time step $t\in[T]$}
        \STATE $q_t\gets{z_t}\,/\,{\|z_t\|_1}$, $p_t^{(j)}\gets{w_t^{(j)}}\,/\,{\|w_t^{(j)}\|_1}$, for all $j\in[J]$;
          \vspace{-3mm}
        \STATE \textbf{output} $p_t \gets \sum_{j=1}^J\, q_{j,\,t}\,p_t^{(j)}$; 
        \STATE \textbf{receive} the reward $r_{a_t,\,t}\in [0,1]$  of selecting the action $a_t\in\calV$ at time step $t$; 
        \FOR {$j\in[J]$}
        \STATE $\Tilde{r}_{i,\,t}^{(j)}\gets 1 - \frac{{\bf 1}(a_t\,=\,i)}{p_{i,\,t}\,+\,\gamma^{(j)}}(1\,-\,r_{a_t,\,t})$, for all $i\in\calV$;
        \STATE $\Tilde{r}_{t}^{(j)}\gets\left[\Tilde{r}_{1,\,t}^{(j)},\dots,\Tilde{r}_{|\calV|,\,t}^{(j)}\right]^\top$;
        \STATE $v_{i,\,t}^{(j)}\gets w_{i,\,t}^{(j)}\,\exp{(\,\eta^{(j)}\,\Tilde{r}_{i,\,t}^{(j)}\,)}$, for all $i\in\calV$;
        \STATE $W_t^{(j)}\gets v_{1,\,t}^{(j)}+\dots+v_{|\calV|,\,t}^{(j)}$;
        \STATE $w_{i,\,t+1}^{(j)}\!\gets\!\beta\,\frac{W_t^{(j)}}{|\calV|}+(1-\beta)\,v_{i,\,t}^{(j)}$, for all $i\in\calV$;
        \STATE $z_{j,\,t+1}\gets z_{j,\,t}\exp{(\,\eta \,\Tilde{r}_t^{(j)\top} \,p_t^{(j)}\,)}$;
		\ENDFOR
		\ENDFOR
	\end{algorithmic}
\end{algorithm}

%% file: alg-BSG.tex
\setlength{\textfloatsep}{3mm}
\begin{algorithm}[t]
	\caption{\mbox{Bandit Sequential Greedy (\banalg).}
	}
	\begin{algorithmic}[1]
		\REQUIRE Time steps $T$ and agents' action sets $\{\mathcal{V}_i\}_{i \myin \calN}$.
		\ENSURE \!Agent actions $\{a_{i,\, t}^\banalg\}_{i\,\in\,\calN}$ at each $t\in[T]$.
		\medskip
		\STATE Each agent $i\in\calN$ initializes an \scenario{EXP3$^\star$-SIX} with the value of the parameters $T$ and $\calV_i$;
		\STATE Denote the \scenario{EXP3$^\star$-SIX} onboard agent $i$ by \expsix{i};
		\STATE Order the agents in $\calN$ such that $\calN=\{1,\dots,|\calN|\}$;
		\FOR {\text{each time step} $t\in [T]$}
 			\FOR {$i=1,\dots,|\calN|$} 
				\STATE \textbf{get} the output $\distfsf_t^{(i)}$ from \expsix{i}; 
				\STATE \textbf{draw} an action $a_{i,\,t}^\banalg$ from the distribution $\distfsf_t^{(i)}$;
			\ENDFOR
			\STATE \textbf{execute} $\{a_{i,\, t}^\banalg\}_{i\,\in\,\calN}$;  
			\STATE $\solbsg_{0,\,t} \gets \emptyset$;
			\FOR {$i = 1, \dots, |\calN|$} 
			\STATE $\solbsg_{i,\,t} \gets \solbsg_{i-1,\,t}\cup \{a_{i,\,t}^\banalg\}$;
			\STATE \textbf{observe} $f_t(\,\solbsg_{i,\,t}\,)$;
			\STATE $r_{i,\, t}^\banalg\gets f_t(\,a_{i,\,t}^\banalg \,|\, \solbsg_{i-1,\,t}\,)$; 
			\STATE \textbf{input} $r_{i,\, t}^\banalg$ to \expsix{i} (per line 8 of \Cref{alg:EXP3star-SIX});
			\ENDFOR
		\ENDFOR
	\end{algorithmic}\label{alg:online}
\end{algorithm}

%% file: 4-Guarantees.tex
\section{Performance Guarantees of \banalg}\label{sec:guarantees}
We present the computational complexity (\Cref{subsec:complexity}) and approximation performance (\Cref{subsec:tracking-regret}) of \banalg. 

\subsection{Computational Complexity of \banalg}\label{subsec:complexity}

\banalg is the first algorithm for \Cref{pr:online} with polynomial computational complexity, quantified below.   

\begin{proposition}[Computational Complexity]\label{pror:computations}
\banalg requires each agent $i\in\calN$ to perform $T$ function evaluations and $O(T\log{T})$ additions and multiplications over $T$ rounds.
\end{proposition}

The proposition holds true since at each $t\in[T]$, \banalg requires each agent $i$ to perform $1$ function evaluation to compute the marginal gain in \banalg's line 14 and $O(\log{T})$ additions and multiplications to run \expsix{i}. 

\begin{remark}[Direct Application of \scenario{EXP3$^\star$-SIX} to \Cref{pr:online} Requires Exponential Running Time in $|\calN|$]
\scenario{EXP3$^\star$-SIX} may be directly applied to \Cref{pr:online}, resulting however an exponential time algorithm since \scenario{EXP3$^\star$-SIX} would then require  $O(T\log{T}\prod_{i\,\in\,\calN}|\calV_i|)$ additions and multiplications.
\end{remark}

\subsection{Approximation Performance of \banalg}\label{subsec:tracking-regret}

We bound \banalg's suboptimality with respect to the optimal actions the agents' would select if they knew the $\{f_t\}_{t\myin[T]}$ a priori. Particularly, we bound \banalg's \textit{tracking regret}, proving that it gracefully degrades with the environment's capacity to select $\{f_t\}_{t\myin[T]}$ adversarially (\Cref{th:half}). 

To present \Cref{th:half}, we first define tracking regret, particularly, \textit{$1/2$-approximate tracking regret} (\Cref{def:tracking_regret}), and then we quantify the environment's capacity to select $\{f_t\}_{t\myin[T]}$ adversarially (\Cref{def:environment_change}). We use the notation:
\begin{itemize}
    \item $\solopt_t\in\arg\max_{a_{i,\,t}\myin\mathcal{V}_{i},\, \forall\, i\myin\calN} \;f_t(\{a_{i,\,t}\}_{i\,\in\, \calN})$ is the optimal actions the agents would select for time step $t$ if they fully knew $f_t$ a priori;
    \item $a_{i,\,t}^\opt$ is agent $i$'s action among the actions in $\solopt_t$;
    \item $\calA_t\triangleq \{a_{i,\,t}\}_{i\,\in\,\calN}$ is the set of all agents' actions at $t$.
\end{itemize}

\begin{definition}[$1/2$-Approximate Tracking Regret]\label{def:tracking_regret}
Consider an arbitrary sequence of action sets $\{\calA_t\}_{t\myin[T]}$. $\{\calA_t\}_{t\myin[T]}$'s $1/2$-approximate tracking regret is\footnote{\Cref{def:tracking_regret} generalizes existing notions of tracking regret \cite{herbster1998tracking,matsuoka2021tracking} to the online submodular coordination \Cref{pr:online}.} 
\vspace{-2mm}
\begin{align}\label{eq:regret}
    &\scenario{Tracking}\text{-}\scenario{Regret}_T^{(1/2)}(\,\{\calA_t\}_{t\myin[T]}\,) \nonumber\\
    &\qquad\qquad\triangleq \frac{1}{2}\sum_{t=1}^{T} f(\,\solopt_t,\,E_t\,) - \sum_{t=1}^{T}f(\,\calA_t,\,E_t\,).
\end{align}
\end{definition}

\Cref{eq:regret} can be further simplified as follows: 
\begin{align}\label{eq:regret_simplified}
    &\scenario{Tracking}\text{-}\scenario{Regret}_T^{(1/2)}(\,\{\calA_t\}_{t\myin[T]}\,) \nonumber\\
    &= \frac{1}{2}\sum_{t=1}^{T} f(\,\solopt_t,\,E_t^{obs}(\solopt_t)\,) - \sum_{t=1}^{T}f(\,\calA_t,E_t^{obs}(\calA_t)\,)\nonumber\\
    &= \frac{1}{2}\sum_{t=1}^{T} f_t(\,\solopt_t\,) - \sum_{t=1}^{T}f_t(\,\calA_t\,),
\end{align}
\noindent per \cref{eq:f_t,eq:partial}.  In more detail,
\cref{eq:regret_simplified} evaluates $\{\calA_t\}_{t\myin[T]}$'s suboptimality against the optimal actions $\{\calA_t^\opt\}_{t\myin[T]}$ the agents would select if they knew the $\{f_t\}_{t\myin[T]}$ a priori.  The optimal total value $\sum_{t=1}^{T} f_t(\solopt_t)$ is discounted by $1/2$ in \cref{eq:regret} since solving exactly \Cref{pr:online} is NP-hard even when $\{f_t\}_{t\myin[T]}$ are known a priori~\cite{sviridenko2017optimal}.  Specifically, the best possible approximation bound in polynomial time is the $1-1/e$~\cite{sviridenko2017optimal}, while the Sequential Greedy algorithm~\cite{fisher1978analysis}, that \banalg extends to the bandit online setting, achieves the near-optimal bound $1/2$. In this paper, 
we prove that \banalg can approximate \textit{Sequential Greedy}'s near-optimal performance by bounding \cref{eq:regret}.

\begin{definition}[Environment's Adversarial Effect \cite{xu2023online}]\label{def:environment_change}
The environment's \emph{adversarial effect} on (i) agent $i$ is  
\begin{equation}
    \Delta_i(T) \triangleq \sum_{t=1}^{T-1} {\bf 1}(\,a_{i,\,t}^\opt\neq a_{i,\;t+1}^\opt\,),
\end{equation}
and 
on (ii) all the agents $\calN$ is
\begin{equation}\label{eq:rate}
	\Delta(T) \triangleq \sum_{i\myin \calN}\Delta_i(T) = \sum_{t=1}^{T-1} \sum_{i\myin \calN} {\bf 1}(\,a_{i,\,t}^\opt\neq a_{i,\;t+1}^\opt\,).
\end{equation}
\end{definition}

$\Delta(T)$ captures the environment's total effect on selecting $\{f_t\}_{t\myin[T]}$ adversarially
by counting how many times the optimal actions of the agents must shift across the $T$ steps to adapt to the changing $f_{t}$.  
The larger the environment capacity  to adversarially select $\{f_t\}_{t\myin[T]}$, the larger the $\Delta(T)$, and the harder for the agents to adapt to near-optimal actions.

\begin{theorem}[Approximation Performance]\label{th:half} \banalg instructs the agents to select actions $\{a_{i,\,t}^\banalg\}_{i\myin\calN,\, t\myin[T]}$ that guarantee
\begin{align}
	&\mathbb{E}\left[\scenario{Tracking}\text{-}\scenario{Regret}_T^{(1/2)}(\,\{a_{i,\,t}^\banalg\}_{i\myin\calN,\, t\myin[T]}\,)\right] \nonumber\\
	&\quad\leq \underbrace{\tilde{O}\left[\sqrt{T|\calN||\bar{\calV}|\left(\Delta(T)+|\calN|\right)}\,\right]}_{{\phi_1}} \nonumber\\
	&\quad \quad+ \underbrace{\log{\left(\frac{1}{\delta}\right)}\;\tilde{O}\left[\sqrt{T|\calN|\sum_{i\myin\calN}\frac{|\calV_i|}{{\Delta}_i(T)+1}} + |\calN|\right]}_{{\phi_2}}\label{eq:th-bound}
\end{align}holds with probability at least $1-\delta$, for any $\delta \in (0,1)$, where the expectation is due to \banalg's internal randomness, $|\bar{\calV}|\triangleq\max_{i\myin\calN}|\calV_i|$, and $\tilde{O}(\cdot)$ hides $\log$~terms.
\end{theorem}

{The proof of \Cref{th:half} is presented in Appendix~\ref{app:BSG}.} \Cref{th:half} bounds the tracking regret of \banalg, as a function of the number of robots, the total time steps $T$, and the environment's total adversarial effect. 
If the environment's total adversarial effect grows slow enough with $T$ such that
{\small\begin{equation}\label{eq:learning-rate}
\tilde{O}\left[\sqrt{\,|\calN|\,T\,|\bar{\calV}|\,\left(\,\Delta(T)+|\calN|\,\right)}\,\right]\Big/T\rightarrow 0 \text{ for $T\rightarrow +\infty$},
\end{equation}}
then \cref{eq:th-bound} implies $f_t(\calA_t)\rightarrow 1/2 f_t(\calA_t^\opt)$ in expectation since then both $\phi_1$ and $\phi_2$ in \cref{eq:th-bound} are sublinear in $T$.\footnote{The term $\phi_2$ in \cref{eq:th-bound} is always sublinear, \ie $\phi_2/T\rightarrow 0$ for $T\rightarrow +\infty$, since $\phi_2$ is bounded by $  \log{\left(\frac{1}{\delta}\right)}\;\tilde{O}\left[|\calN|\left(\sqrt{T|\,\bar{\calV}|} + 1\right)\right]$.}
In other words, when \cref{eq:learning-rate} holds true, then \banalg enables the agents to asymptotically learn (adapt) to coordinate as if they knew $f_1,\dots,f_{T}$ a priori, matching the performance of the near-optimal \sg. 
For example, \cref{eq:learning-rate} holds true in environments whose evolution is unknown yet predefined, instead of being adaptive to the agents' actions.  Then, $\Delta(T)$ is uniformly bounded since increasing the discretization density of time horizon $H$, \ie increasing the number of time steps $T$, does not affect the environment's evolution.  Thus,
$\Delta(T)/T\rightarrow 0$ for $T\rightarrow +\infty$, which implies  \cref{eq:learning-rate}. The result agrees with the intuition that the agents should be able to adapt to an unknown but non-adversarial environment when they {re-select actions with high enough frequency.}
\begin{figure}[b!]
    \captionsetup{font=footnotesize}
    \centering
    \includegraphics[width=0.61\columnwidth]{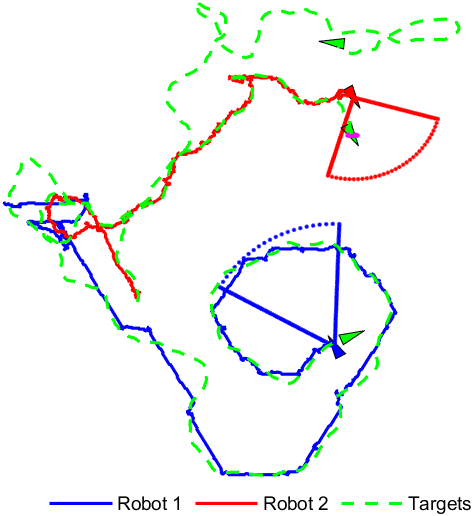}\\
    \caption{\textbf{Target-Tracking Instance of $2$ Robots Tracking $3$ Targets.} Both the robots and the targets move on the same plane. Each robot can collect range and bearing measurements of a target but only if the target is inside the robot's field of view.
    The measurements are assumed corrupted with zero-mean Gaussian noise, with increasing variance the further away the target is from the robot.
    The targets' motion model is unknown to the robots, thus the robots coordinate based only on past observations. 
    }\label{fig:partial-sim}
\end{figure}

%% file: 5-Experiments.tex
\section{Numerical Evaluation in Multi-Target Tracking Tasks with Multiple Robots}\label{sec:experiments}

We evaluate \banalg in simulated scenarios of target tracking with multiple robots, where the robots carry noisy sensors with limited field of view to observe the targets. 
We consider scenarios where $2$ robots pursue $2$, $3$, or $4$ targets.  For each scenario, we first consider non-adversarial targets and, then, adversarial targets: the non-adversarial targets traverse predefined trajectories, independently of the robots' motion; whereas, the adversarial targets maneuver in response to the robots' motion.  \textit{In both cases, the targets' future motion and maneuvering capacity are unknown to the robots.} 

Particularly, we first evaluate \banalg's effectiveness at different action-selection frequencies (10, 20, 50, and 100Hz). To this end, we consider scenarios of 2 robots pursuing 2 non-adversarial targets in \Cref{subsec:frequency}, validating the theoretical results in \Cref{sec:guarantees}. Then, we evaluate \banalg's effectiveness in enabling the robots to pursue the targets. To this end, we consider scenarios where $2$ robots pursue $2$, $3$, or $4$ targets, first focusing on non-adversarial targets (\Cref{subsec:bsg_non_adversarial}) and, then, on adversarial targets  (\Cref{subsec:bsg_adversarial}).  We also compare \banalg's performance with a greedy heuristic, showcasing \banalg's superiority. We provide video demonstrations for all simulation scenarios at \href{https://bit.ly/3WlxcUy}{\red{https://bit.ly/3WlxcUy}}.

\vspace{1mm}
\myParagraph{Common Simulation Setup across Simulated Scenarios} 

\paragraph{Targets} The targets move on a 2D plane.   We introduce the targets' motion model within each particular scenario considered in \Cref{subsec:frequency} and \Cref{subsec:bsg_adversarial}.

Henceforth, $\calT$ denotes the set of targets.

\paragraph{Robots}
The robots move in the same 2D environment as the targets. To move in the environment, each robot $i\in\calN$ can perform one of the actions  $\calV_i\triangleq$ \{``upward'', ``downward'', ``left'', ``right'', ``upleft'', ``upright'', ``downleft'', ``downright''\}  at a constant speed.

\paragraph{Sensing} 
We consider that each robot $i$ has a range and bearing sensor to collect measurements about the targets' position inside its field of view. After selecting actions $\{a_{i,\,t}\}_{i\myin\calN}$ at time $t$, the robots share their measurements with one another, enabling each robot $i$ to have an estimate of $d_t(a_{i,\,t},\,j)$, the distance from robot $i$ to target $j$, given that $j$ is observed by a robot as a result of actions $\{a_{i,\,t}\}_{i\myin\calN}$.

\paragraph{Objective Function} The robots coordinate their actions $\{a_{i,\,t}\}_{i\myin \calN}$ to maximize at each time step $t$\footnote{The objective function in \cref{eq:distance} is a non-decreasing and submodular function.  The proof is presented in the Appendix.}

\vspace{-2mm}
{\small
\begin{equation}\label{eq:distance}
   f_t(\,\{a_{i,\,t}\}_{i\myin \calN}\,)\,=\, \sum_{j\myin\calT}\left[-\sum_{i\myin \calN_j} \; \frac{1}{d_t(a_{i,\,t},\,j)}\right]^{-1},
\end{equation}}where $\calN_j$ is the set of robots that can observe the target $j$, $d_t(a_{i,\,t},\,j)$ is the distance between robot $i$ and the estimated location  of target $j$.  Therefore, {$d_t(a_{i,\,t},\,j)$} is known only once robot $i$ has executed its action $a_{i,\,t}$ and the location estimate of target $j$ at time step $t$ has been computed. {In particular, it is assumed that the total number of targets in the environment is known to the robots such that \Cref{ass:partial} is satisfied.}

By maximizing $f_t$, the robots aim to collaboratively keep the targets inside their field of view.  For example, when \underline{no} robot has target $j$ inside its field of view, \ie when $\calN_j=\emptyset$, which is equivalent to target $j$ being infinitely far away from all robots, it is $\left[-\sum_{i\myin \calN_j} \; {1}/{d_t(a_{i,\,t},\,j)}\right]^{-1}=-\infty$ ---to account for the feasibility of our implementation, when $\calN_j=\emptyset$, we set $\left[-\sum_{i\myin \calN_j} \; {1}/{d_t(a_{i,\,t},\,j)}\right]^{-1}=-4d_{max}$, where $d_{max}$ is the largest sensing range among the robots.  On the other end of the spectrum, when a robot $i$ achieves $0$ estimated distance from a target $j$, \ie when $d_t(a_{i,t},j)=0$, then indeed $\left[-\sum_{i\myin \calN_j} {1}/{d_t(a_{i,t},j)}\right]^{-1}=0$.

\begin{figure}[t!]
  \captionsetup{font=footnotesize}
  \centering
  \includegraphics[width=.9\columnwidth]{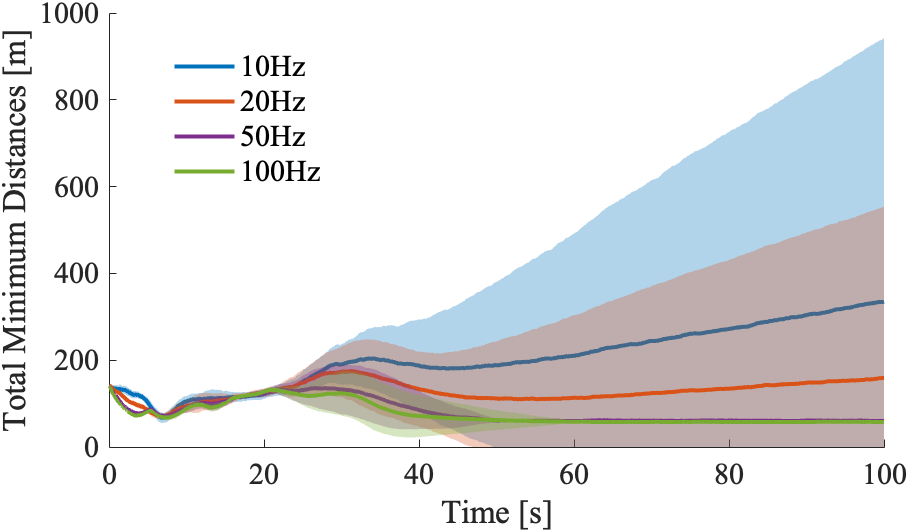}\\
  \caption{\textbf{{\smaller \sf BSG}'s Performance for Various Action-Selection Frequencies.} Four  frequencies are depicted, in a target tracking scenario where $2$ robots pursue $3$ non-adversarial targets; the targets predefined trajectories are shown in Fig.~\ref{fig:non_adversarial}(d).  The results are averaged over $50$ Monte-Carlo trials.
}\label{fig:frequencies}
\end{figure}

\paragraph{Performance Metric}
To measure how closely the robots track the targets, we consider a 
\textit{total minimum distance} metric.  We define the metric as the sum of the distances between each target and its nearest robot, {whether this target is observed by any robot or not}. 

\myParagraph{Computer System Specifications} We ran all simulations in MATLAB 2022b on a Windows laptop equipped with the Intel Core i7-10750H CPU @ 2.60 GHz and 16 GB RAM. 

\textbf{Code.} {Our code is available at:} \href{https://github.com/UM-iRaL/bandit-sequential-greedy}{\red{https://github.com/UM-iRaL/bandit-sequential-greedy}}.

\input{5-fig-non-adversarial}

\subsection{Evaluation of \banalg at Various Reaction Frequencies}\label{subsec:frequency}

We evaluate the capacity of \banalg to improve its performance when the robots’ action-selection frequency increases. Particularly, we test \banalg when the robots’ action-selection frequency increases from 10Hz to 20Hz to 50Hz to 100Hz, in a scenario where 2 robots pursue 3 non-adversarial targets whose predefined trajectories are shown in Fig.~\ref{fig:non_adversarial}(d).

\myParagraph{Results} The simulation results are presented in Fig.~\ref{fig:frequencies}, averaged across $50$ Monte-Carlo trials.  They validate the analysis in \Cref{subsec:tracking-regret}, specifically, that the higher is the action-selection frequency the better \banalg learns and, thus, the closer the robots pursue the targets.  Particularly, although at $10$Hz, the robots fail to ``learn'' the targets' future motion, failing to reduce their distance to them, the  situation improves at $20$Hz, and even further at $50$Hz and $100$Hz. In the latter two cases, the robots closely track the targets, maintaining on average a non-increasing distance to them, proportional to the field of view of the robots: the field of view of the robots has a radius of $150$m, and the achieved total minimum distance at $50$Hz and $100$Hz is less than $100$m.

\input{5-fig-adversarial}

\subsection{Evaluation of \banalg in Non-Adversarial Target Tracking}\label{subsec:bsg_non_adversarial}

We evaluate \banalg in simulated target tracking scenarios where the targets are non-adversarial, \ie they traverse predefined trajectories that are non-adaptive to the robots' locations. To this end, we first describe a heuristic baseline against which we compare \banalg and the simulation setup.  

\myParagraph{Benchmark Algorithm} 
We compare \banalg with a heuristic version of the Sequential Greedy that selects actions at each $t$ based on the previous $f_{t-1}$.   We denote the algorithm by \sgh.  \sgh \xspace selects actions per the rule:
\begin{align}\label{eq:sgd}
  &a_{i,\,t}^{{\sgh}} \,\\\nonumber
  &\quad\in\, \max_{a\myin\calV_i}\;
  f_{t-1}(\,a\;|\;\{a_{1,\,t}^{\sgh},\ldots,a_{i-1,\,t}^{\sgh}\}\,).
\end{align}

\myParagraph{Simulation Setup} 
We consider three scenarios of non-adversarial target tracking: 
(i) $2$ robots vs.~$2$ targets, where the targets traverse straight lines with a crossing (Fig.~\ref{fig:non_adversarial}(a)--(c)); (ii) $2$ robots vs.~$3$ targets, where the targets traverse straight lines and circles with a crossing (Fig.~\ref{fig:non_adversarial}(d)--(f)); and (iii) $2$ robots vs.~$4$ targets, where the targets diverse and traverse straight lines with turns (Fig.~\ref{fig:non_adversarial}(g)--(i)).  Each robot and target have different speeds, but we assume that all targets move with less speed than the robots.
In all scenarios, the robots re-select actions with frequency $20$Hz. We evaluate the algorithms across $50$ Monte-Carlo trials.

\vspace{-0mm}
\myParagraph{Results} The simulation results are presented in Fig.~\ref{fig:non_adversarial}.  
The following observations are due:
(i)~~\banalg outperforms \sgh\xspace in all scenarios (Fig.~\ref{fig:non_adversarial}(c)(f)(i)).  
{In all the cases of $2$ robots vs.~$2$ targets (Fig.~\ref{fig:non_adversarial}(a)--(c)), $2$ robots vs.~$3$ targets (Fig.~\ref{fig:non_adversarial}(d)--(f)), and of $2$ robots vs.~$4$ targets (Fig.~\ref{fig:non_adversarial}(g)--(i)), \banalg maintains near-constant distances to the targets.} 
(ii)~~\banalg enables collaborative behaviors such as robots switching targets to improve speed compatibility.
Particularly, in all Fig.~\ref{fig:non_adversarial}(a)(d)(g), we observe that the robots eventually switch their corresponding targets to chase.  The reason is that the faster robots can match the faster targets. \textit{This desirable switching behavior {may} emerge even though the robots are unaware of the targets' speed and overall motion model.} 
{(iii)~~\sgh\xspace instructs the robots to chase only one group of targets once the targets disperse. This is because \sgh\xspace is based \textit{merely} on the outdated detected target locations of the last time step. But looking at only the last time step can be misleading. 
For example, if a robot loses all targets at a time step, then it will have nothing to feed into the \sgh\xspace in the next time step and, thus, it will then start to randomly scan all of its actions until it tracks some group of targets again. Then, the robot will keep tracking the new targets and may never get back to the past ones. In contrast, \banalg uses the reward information of all past selected actions to predict the best actions for the robots, such that a robot can track the same targets again even after it has lost them.}

\subsection{Evaluation of \banalg in Adversarial Target Tracking}\label{subsec:bsg_adversarial}

We evaluate \banalg in simulated target tracking scenarios where the targets are adversarial, \ie they traverse predefined trajectories that are adaptive to the robots' locations. 

\myParagraph{Simulation Setup} The setup is the same as in \Cref{subsec:bsg_non_adversarial} with the exception that here the targets adapt their motion to the robots' motion: as long as all robots are more than $50$m away from a target, the target performs a random walk; but if any robot is within $50$m from a target, then this target increases its speed by $10$m/s for $5$s, pointing it to a direction that maximizes the average distance from all robots.

\myParagraph{\textbf{Results}}
The simulation results are shown in Fig.~\ref{fig:adversarial}. Similarly to the non-adversarial case, (i) \banalg outperforms \sgh\xspace across all scenarios (Fig.~\ref{fig:adversarial}(c)(f)(i)), and (ii) \banalg enables collaborative behaviors among the robots, where fast robots that originally track slow targets eventually switch to faster targets, and slow robots that originally track fast targets switch to slower targets (see, \eg Fig.~\ref{fig:adversarial}(a)).

%% file: 5-fig-non-adversarial.tex
\newcommand{\introFigTitleWidth}{0.2cm}
\newcommand{\introFigColWidth}{5cm}
\newcommand{\introFigSpacing}{\hspace{-2mm}}
\newcommand{\intoFigNameSpacing}{}
\newcommand{\advFigColWidth}{6cm}

\begin{figure*}[t!]
    \captionsetup{font=footnotesize}
	\begin{center}
	\hspace{-9cm}
      \begin{minipage}{\columnwidth}
            \begin{tabular}{p{\introFigTitleWidth}p{\introFigColWidth}p{\introFigColWidth}p{\advFigColWidth}}%
            \begin{minipage}{\introFigTitleWidth}%
            \end{minipage}
            &            
            \begin{minipage}{\introFigColWidth}%
                  \centering
                  \rotatebox{0}{{\sf \smaller\textbf{BSG}}\vspace{-4cm}}
            \end{minipage}
            &            
            \begin{minipage}{\introFigColWidth}%
                  \centering
                  \rotatebox{0}{{\sf \smaller\textbf{SG-Heuristic}}\vspace{-4cm}}
            \end{minipage}
            &
            \begin{minipage}{\advFigColWidth}%
                  \centering
                  \rotatebox{0}{{\sf \smaller\textbf{BSG}} \textbf{vs.~}{\sf \smaller\textbf{SG-Heuristic}} \vspace{-4cm}}
            \end{minipage}
            \\
            \begin{minipage}{\introFigTitleWidth}%
                  \rotatebox{90}{\hspace{.6cm}\textbf{2 vs.~2} \vspace{-4cm}}
            \end{minipage}
            &
            \begin{minipage}{\introFigColWidth}%
                  \vspace{3mm}
                  \centering%
                  \includegraphics[width=\columnwidth]{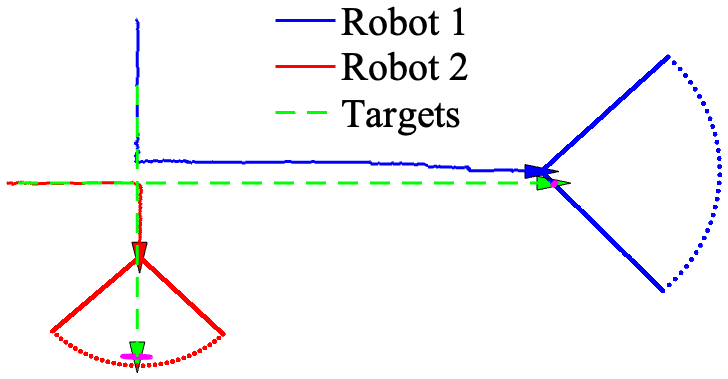} \\
                  \vspace{2.2mm}
                  \caption*{(a) {\smaller \sf BSG}: 2 robots and 2 targets.
                  }
            \end{minipage}
            &
            \begin{minipage}{\introFigColWidth}
                  \vspace{1.5mm}
                  \centering%
                  \includegraphics[width=.66\columnwidth]{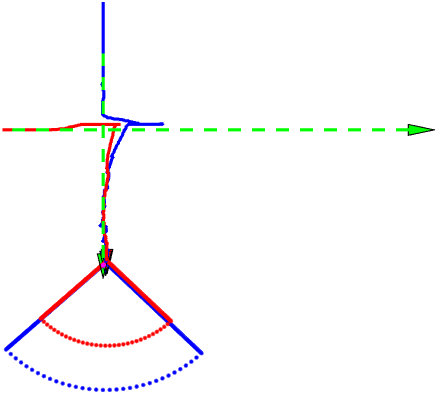} \\ 
                  \caption*{(b) {\smaller \sf SG\text{-}Heuristic}: 2 robots and 2 targets.
                  }
            \end{minipage}
            &
            \begin{minipage}{\advFigColWidth}%
                  \vspace{2mm}
                  \centering%
                  \hspace*{.4mm}
                  \includegraphics[width=.97\columnwidth]{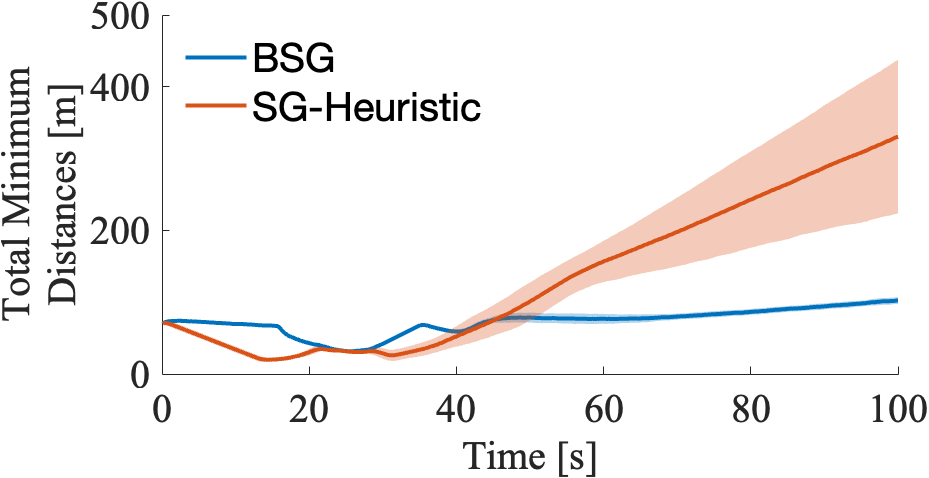} \\
                  \vspace{-0.7mm}
                  \caption*{
                  (c) {\smaller \sf BSG} vs. {\smaller \sf SG\text{-}Heuristic}:
                  2 robots and 2 targets.
                  }
            \end{minipage}
            \\
            \begin{minipage}{0.3cm}%
                  \rotatebox{90}{\hspace{-4mm}\textbf{2 vs.~3} \hspace{-1.8cm}}
            \end{minipage}
            &
            \begin{minipage}{\introFigColWidth}%
                  \vspace{5mm}
                  \centering%
                  \includegraphics[width=\columnwidth]{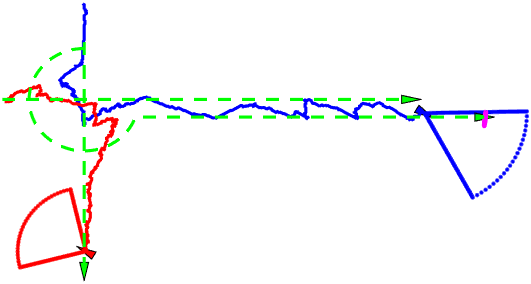} \\
                  \vspace{1.4mm}
                  \caption*{(d) {\smaller \sf BSG}: 2 robots and 3 targets.
                  }
            \end{minipage}
            &
            \begin{minipage}{\introFigColWidth}
                  \vspace{2mm}
                  \centering%
                  \includegraphics[width=.9\columnwidth]{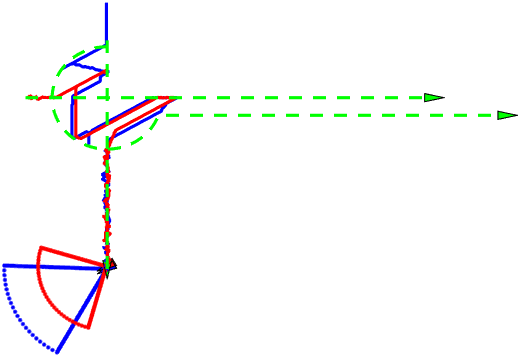} \\ 
                  \vspace{0mm}
                  \caption*{(e) {\smaller \sf SG\text{-}Heuristic}: 2 robots and 3 targets. 
            }
            \end{minipage}
            &
            \begin{minipage}{\advFigColWidth}%
                  \vspace{4mm}
                  \centering%
                  \includegraphics[width=.97\columnwidth]{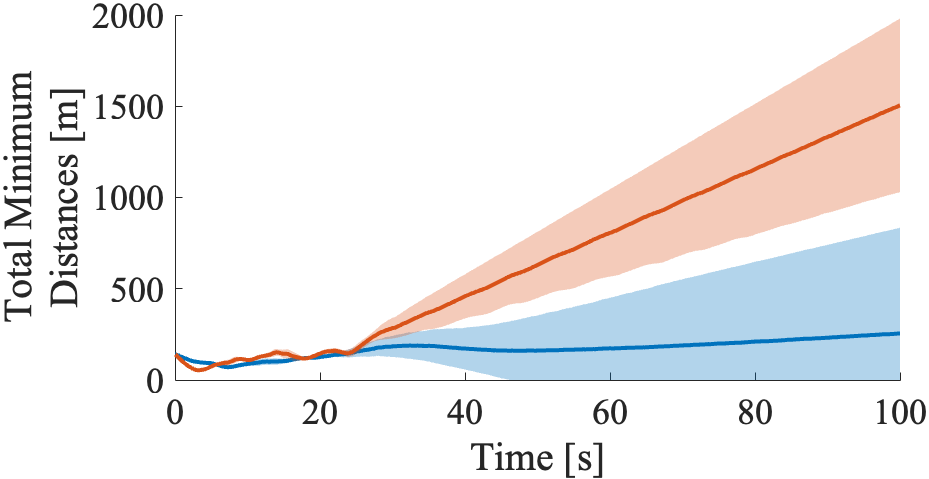} \\
                  \vspace{-1mm}
                  \caption*{(f) {\smaller \sf BSG} vs. {\smaller \sf SG\text{-}Heuristic}: 2 robots and 3 targets.
            }
            \end{minipage}\\
            \begin{minipage}{0.3cm}%
                  \rotatebox{90}{\hspace{-3mm}\textbf{2 vs.~4}\hspace{-2cm}}
            \end{minipage}
            &
            \begin{minipage}{\introFigColWidth}%
            \vspace{2mm}
            \centering%
            \includegraphics[width=.8\columnwidth]{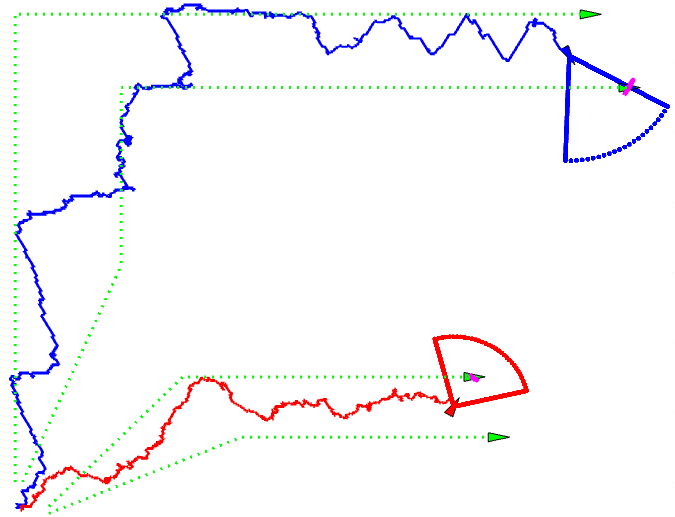} \\
            \vspace{0.5mm}
            \caption*{(g) {\smaller \sf BSG}: 2 robots and 4 targets.
            }
            \end{minipage}
            &
            \begin{minipage}{\introFigColWidth}
                  \centering%
            \vspace{2mm}
            \includegraphics[width=.79\columnwidth]{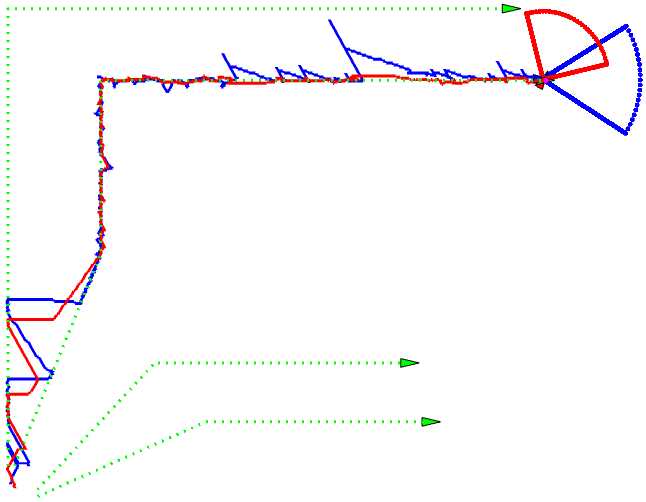} \\ 
             \vspace{0.2mm}
            \caption*{(h) {\smaller \sf SG\text{-}Heuristic}: 2 robots and 4 targets.
            }
            \end{minipage}
            &
            \begin{minipage}{\advFigColWidth}%
            \vspace{2mm}
            \centering%
            \vspace{1mm}
            \includegraphics[width=\columnwidth]{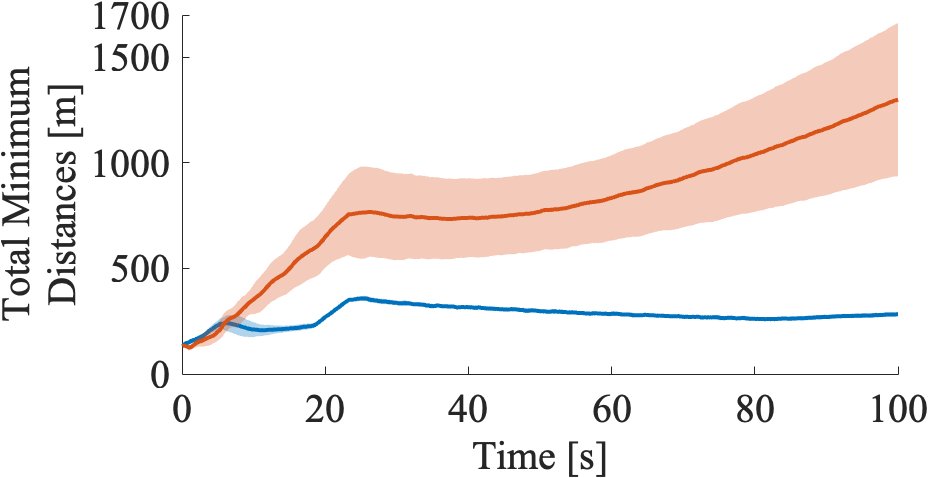} \\
            \vspace{-1mm}
            \caption*{(i) {\smaller \sf BSG} vs. {\smaller \sf SG\text{-}Heuristic}:
            2 robots and 4 targets.
            }
            \end{minipage}
            \end{tabular}
	\end{minipage} 
	\caption{\textbf{Non-Adversarial Target Tracking with Multiple Robots: $2$ Robots Pursuing $2$, $3$, or $4$ Targets.}   
	The robots select actions either per {\smaller \sf BSG}, or per the greedy heuristic {\smaller \sf SG\text{-}Heuristic}, re-selecting actions with frequency $20$Hz. Across the two algorithm cases, the targets traverse the same predefined trajectories, which are non-adaptive to the robots' motion. (a),(d),(g): The robots use {\smaller \sf BSG} against $2$, $3$, and $4$ targets, respectively; (b),(e),(h): the robots use {{\smaller \sf SG\text{-}Heuristic}} against $2$, $3$, and $4$ targets, respectively. (c),(f),(i): Comparison of {\smaller \sf BSG}'s and {{\smaller \sf SG\text{-}Heuristic}}'s  \mbox{average effectiveness over $50$ Monte-Carlo trials.}
	}\label{fig:non_adversarial}
	\vspace{-7mm}
	\end{center}
\end{figure*}

%% file: 5-fig-adversarial.tex
\begin{figure*}[t!]
    \captionsetup{font=footnotesize}
	\begin{center}
	\hspace{-9cm}
    \begin{minipage}{\columnwidth}
        \begin{tabular}{p{\introFigTitleWidth}p{\introFigColWidth}p{\introFigColWidth}p{\advFigColWidth}}%
        \begin{minipage}{\introFigTitleWidth}%
        \end{minipage}
        &            
        \begin{minipage}{\introFigColWidth}%
              \centering
              \rotatebox{0}{\sf \smaller\textbf{BSG}\vspace{-4cm}}
        \end{minipage}
        &            
        \begin{minipage}{\introFigColWidth}%
              \centering
              \rotatebox{0}{{\sf \smaller\textbf{SG-Heuristic}}\vspace{-4cm}}
        \end{minipage}
        &
        \begin{minipage}{\advFigColWidth}%
              \centering
              \rotatebox{0}{{\sf \smaller\textbf{BSG}} \textbf{\small vs.~}{\sf \smaller\textbf{SG-Heuristic}}\vspace{-4cm}}
        \end{minipage}
        \\
        \begin{minipage}{\introFigTitleWidth}%
              \rotatebox{90}{\hspace{.5cm}\textbf{2 vs.~2}\vspace{-4cm}}
        \end{minipage}
        &
        \begin{minipage}{\introFigColWidth}%
            \vspace{5mm}
            \centering%
            \includegraphics[width=.6\columnwidth]{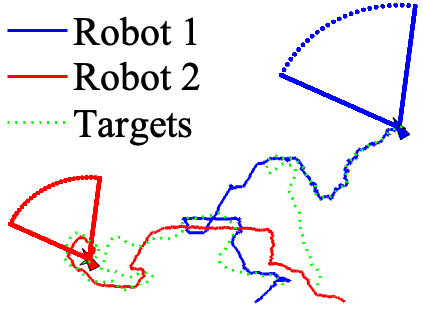} \\
            \vspace{4mm}
            \caption*{(a) {\smaller \sf BSG}: 2 robots and 2 targets.
            }
        \end{minipage}
        &
        \begin{minipage}{\introFigColWidth}
              \vspace{6.3mm}
              \centering%
              \includegraphics[width=.7\columnwidth]{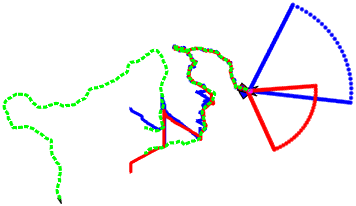} \\ 
              \vspace{4.8mm}
              \caption*{(b) {\smaller \sf SG\text{-}Heuristic}: 2 robots and 2 targets.
              }
        \end{minipage}
        &
        \begin{minipage}{\advFigColWidth}%
              \vspace{2.3mm}
              \centering%
              \hspace*{.4mm}
              \includegraphics[width=.97\columnwidth]{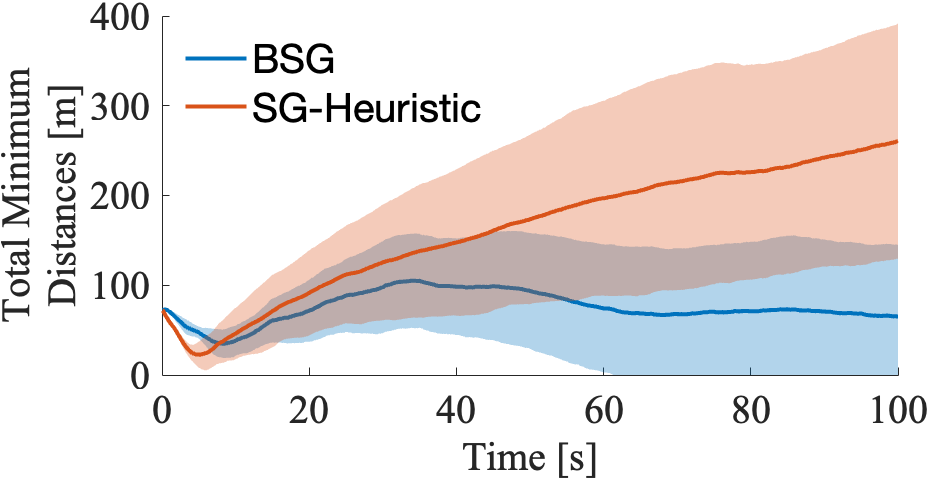} \\
              \vspace{-0.8mm}
              \caption*{
              (c) {\smaller \sf BSG} vs. {\smaller \sf SG\text{-}Heuristic}:
              2 robots and 2 targets.
              }
        \end{minipage}
        \\
        \begin{minipage}{0.3cm}%
              \rotatebox{90}{\hspace{-.3cm}\textbf{2 vs.~3}\hspace{-1.8cm}}
        \end{minipage}
        &
        \begin{minipage}{\introFigColWidth}%
              \vspace{2mm}
              \centering%
              \includegraphics[width=.7\columnwidth]{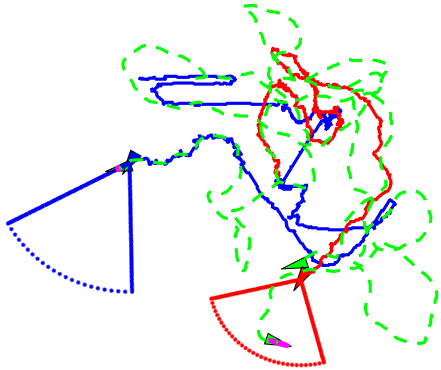} \\
              \vspace{3mm}
              \caption*{(d) {\smaller \sf BSG}: 2 robots and 3 targets.
              }
        \end{minipage}
        &
        \begin{minipage}{\introFigColWidth}
        \vspace{8mm}
              \centering%
        \includegraphics[width=.7\columnwidth]{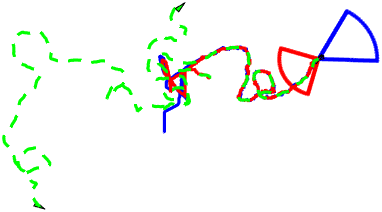} \\ 
         \vspace{6.6mm}
        \caption*{(e) {\smaller \sf SG\text{-}Heuristic}: 2 robots and 3 targets. 
        }
        \end{minipage}
        &
        \begin{minipage}{\advFigColWidth}%
        \vspace{4mm}
        \centering%
        \includegraphics[width=\columnwidth]{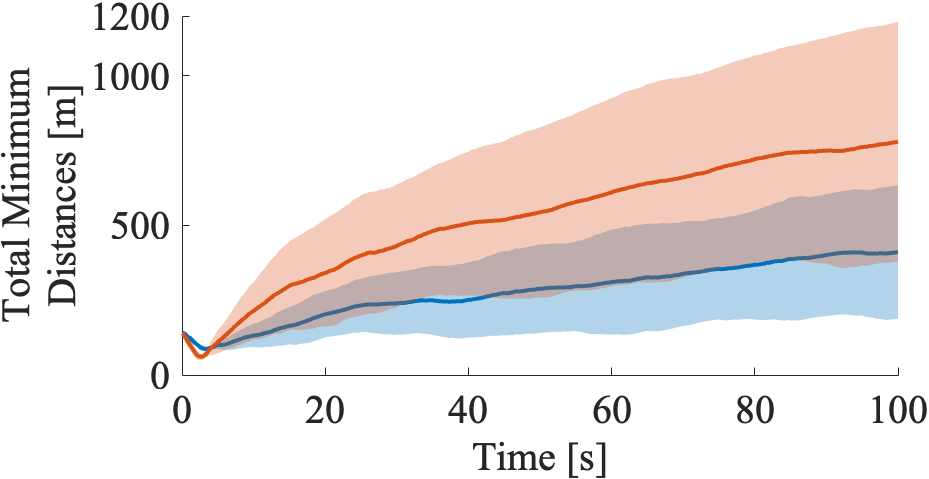} \\
        \vspace{-0.8mm}
        \caption*{(f) {\smaller \sf BSG} vs. {\smaller \sf SG\text{-}Heuristic}:
        2 robots and 3 targets.
        }
        \end{minipage}\\
        \begin{minipage}{0.3cm}%
              \rotatebox{90}{\hspace{-.4cm}\textbf{2 vs.~4}\hspace{-2cm}}
        \end{minipage}
        &
        \begin{minipage}{\introFigColWidth}%
        \vspace{3mm}
        \centering%
        \includegraphics[width=\columnwidth]{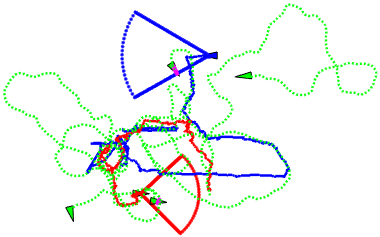} \\
        \vspace{0.2mm}
        \caption*{(g) {\smaller \sf BSG}: 2 robots and 4 targets.
        }
        \end{minipage}
        &
        \begin{minipage}{\introFigColWidth}
        \vspace{4mm}
              \centering%
        \includegraphics[width=\columnwidth]{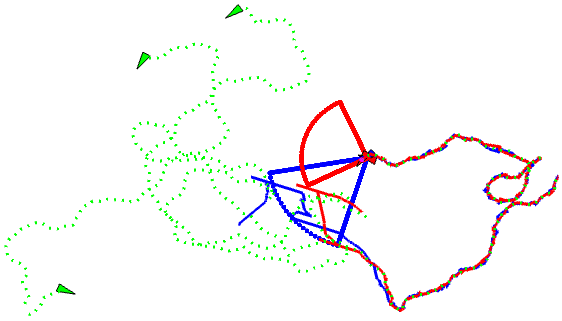} \\ 
        \vspace{1.5mm}
        \caption*{(h) {\smaller \sf SG\text{-}Heuristic}: 2 robots and 4 targets.
        }
        \end{minipage}
        &
        \begin{minipage}{\advFigColWidth}%
        \vspace{4mm}
        \centering%
        \includegraphics[width=\columnwidth]{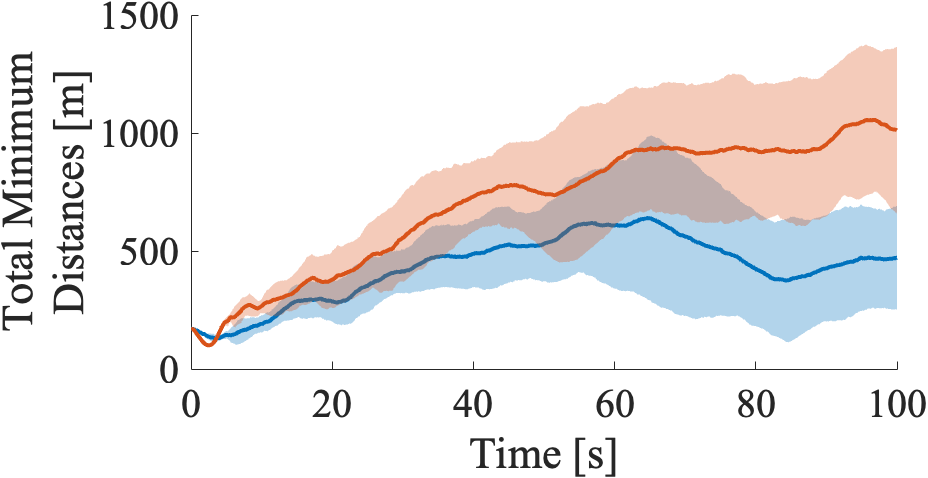} \\
         \vspace{-5mm}
        \hspace{-1mm}
        \caption*{(i) {\smaller \sf BSG} vs. {\smaller \sf SG\text{-}Heuristic}:
        2 robots and 4 targets.
        }
        \end{minipage}
        \end{tabular}
\end{minipage} 
	\caption{\textbf{Adversarial Target Tracking with Multiple Robots: $2$ Robots Pursuing $2$, $3$, or $4$ Targets.} 
	The robots select actions either per {\smaller \sf BSG}, or per the greedy heuristic {\smaller \sf SG\text{-}Heuristic}, re-selecting actions with frequency $20$Hz.  The targets adapt their motion to the robots' motion: as long as all robots are more than $50$m away from a target, the target performs a random walk; but if any robot is within $50$m from a target, then this target increases its speed by $10$m/s for $5$s, pointing it to a direction that maximizes the average distance from all robots. (a),(d),(g): The robots use {\smaller \sf BSG} against $2$, $3$, and $4$ targets, respectively; (b),(e),(h): the robots use {{\smaller \sf SG\text{-}Heuristic}} against $2$, $3$, and $4$ targets, respectively. (c),(f),(i): Comparison of {\smaller \sf BSG}'s and {{\smaller \sf SG\text{-}Heuristic}}'s  \mbox{average effectiveness over $50$ Monte-Carlo trials.}
	}\label{fig:adversarial}
	\vspace{-9mm}
	\end{center}
\end{figure*}

%% file: 6-Conclusion.tex
\section{Conclusion} \label{sec:con}
\myParagraph{Summary} 
We introduced the first algorithm for online submodular coordination in unpredictable and partially observable environments with bandit feedback. Particularly, \banalg is the first polynomial time algorithm with bounded tracking regret for \Cref{pr:online}, requiring only one function evaluation and $O(\log{T})$ additions and multiplications per agent per time step.
The tracking regret bound gracefully degrades with the environments' capacity to change, quantifying how frequently the agents should re-select actions to learn to coordinate as if they fully knew the future a priori.  
\banalg generalizes the seminal {Sequential Greedy} algorithm~\cite{fisher1978analysis} to \Cref{pr:online}'s bandit setting. To this end, we first provided the \scenario{EXP3$^\star$-SIX} algorithm for the problem of \textit{tracking the best action with bandit feedback}. Then, using \scenario{EXP3$^\star$-SIX} as a subroutine, we proposed the \banalg algorithm for \Cref{pr:online}, leveraging submodularity, inspired by the algorithm in~\cite{xu2023online}.
We validated \banalg in simulated scenarios of target tracking with multiple robots, demonstrating how \banalg can enable the robots to collaborate and adapt. 

\myParagraph{Limitations} 
\banalg has the main limitations: (i) \banalg is a centralized algorithm where each robot needs to know actions selected by all previous robots to make a decision (\Cref{alg:online}); (ii) \banalg requires a fine enough time discretization to achieve a near-optimal performance (Fig.~\ref{fig:frequencies}); and (iii)  
\banalg can have $O(T)$ tracking regret in the worst case since the environment can arbitrarily evolve such that $\Delta(T)$ is $O(T)$; this is a fundamental limit that emerges even in the single-agent case of the \textit{tracking the best expert} problem~\cite[Chapter 11]{lattimore2020bandit}, due to the challenging unpredictable environment (\Cref{pr:tracking-the-best-action}).  To overcome this fundamental limit, we will leverage external advice about the evolution of the environment, managing the risk of erroneous advice,
as discussed next.

\myParagraph{Future Work: Leveraging External Advice} 
\banalg selects actions assuming that the environment may evolve arbitrarily in the future.  This assumption is pessimistic when there is side information about the environment's evolution.  
We will extend \banalg such that it can leverage side information in the form of external advice, \eg {in the form of external commands originated by human operators or machine learning algorithms}. 
We will guarantee that the algorithm is \textit{consistent} and \textit{robust}:  (i)~~\textit{consistent}: the algorithm will guarantee enhanced performance when the external advice is better than \banalg in hindsight; (ii)~~\textit{robust}: but when the advice is poor (worse than \banalg), the algorithm will still guarantee a comparable performance to the \banalg algorithm.

%% file: App-1.tex
\section{Proof of \Cref{th:EXP3star-SIX}}\label{app:EXP3star-SIX}

\scenario{EXP3$^\star$-SIX}'s regret  can be decomposed into two parts, as follows:
\begin{equation}\label{aux1:1}
    \sum_{t=1}^{T} \left(r_{a^\star,\,t} - r_{t}^{\top}p_t\right) = \sum_{t=1}^{T} \left(r_{t}^{\top}\distfsf_t^{(j)} - r_{t}^{\top}p_t\right) + \sum_{t=1}^{T} \left(r_{a^\star,\,t} - r_{t}^{\top}\distfsf_t^{(j)}\right).
\end{equation}

It suffices to prove that \cref{aux1:1} is bounded by $$4\sqrt{2T\Big(\Bar{P}(T)|\calV|\log{(|\calV|T)} + \log{(1+\log{T})\Big)}} + \left(\sqrt{\frac{|\calV|T}{\Bar{P}(T)\log{(|\calV|T)}}} + 1\right)\log{\left(\frac{1}{\delta}\right)},$$

where $\Bar{P}(T)\triangleq P(T)+1$. 

To this end, we first consider the following special cases:
\begin{itemize}[leftmargin=3mm]
    \item if $|\calV|\;=1$, then $\sum_{t=1}^{T} \left(r_{a^\star,\,t} - r_{t}^{\top}\distfsf_t^{(j)}\right)=0$ and, thus,  \Cref{th:EXP3star-SIX} holds true;
    \item if $T=1$ and $|\calV|\;\geq 2$, then  $\sum_{t=1}^{T} \left(r_{a^\star,\,t} - r_{t}^{\top}\distfsf_t^{(j)}\right)\leq 1$ since $r_t\in[0,1]^{|\calV|}$.  Since for $T=1$ it also is $\Bar{P}(T)=1$ and, as a result, $4\sqrt{2T(\Bar{P}(T)|\calV|\log{(|\calV|T)} + \log{(1+\log{T}))}}\geq 8\sqrt{\log{2}} > 2$,  \Cref{th:EXP3star-SIX} again holds true; 
    \item if $T=2$ and $|\calV|\;\geq 2$, then similarly to above $\sum_{t=1}^{T} \left(r_{a^\star,\,t} - r_{t}^{\top}\distfsf_t^{(j)}\right)\leq 2$, $\Bar{P}(T)$ is equal to either $1$ or $2$, and $$4\sqrt{2T(\Bar{P}(T)|\calV|\log{(|\calV|T)} + \log{(1+\log{T}))}}\geq 8\sqrt{4\log{2}+\log{(1+\log{2})}} > 2.$$
    Thereby, \Cref{th:EXP3star-SIX} still holds true. 
\end{itemize}

We now consider the last case where  $T\geq3$ and $|\calV|\;\geq 2$. For this case, we start with bounding the first part of \cref{aux1:1}.
From \cite[Theorem 2.2]{cesa2006prediction}, we have
\begin{equation}\label{aux1:MWU_bound}
    \sum_{t=1}^{T} \left(r_{t}^{\top}\distfsf_t^{(j)} - r_{t}^{\top}p_t \right)\leq 2\eta T+\frac{\log{J}}{\eta} = 2\sqrt{2T\log{J}},
\end{equation}where we choose $\eta=\sqrt{\log{J}/(2T)}$.

We next bound the second part of \cref{aux1:1}.
To this end, from \cite[Appendix B.2]{neu2015explore}, we have that
\begin{align}
    \hspace{-4mm}\sum_{t=1}^{T} \left(r_{a^\star,\,t} - r_{t}^{\top}\distfsf_t^{(j)} \right) \leq &\;\frac{\Bar{P}(T)\log{|\calV|}}{\eta} + \frac{1}{\eta}\log{\frac{1}{\beta^{P(T)}(1-\beta)^{T-P(T)-1}}}+ \frac{\Bar{P}(T)\log{|\calV|}+P(T)\log{\left(\frac{eT}{P(T)}\right)}+\log{\left(\frac{1}{\delta}\right)}}{2\gamma} \nonumber\\
    &+ \left(\frac{\eta}{2}+\gamma\right)|\calV|T + \left(\frac{\eta}{2}+\gamma\right)\frac{\log{\left(\frac{1}{\delta}\right)}}{2\gamma}
\end{align}holds true with probability at least $1-\delta$, $\delta\in(0,1)$. Choosing now $\beta=\frac{1}{T-1}$ and $\gamma^{(j)}=\frac{1}{2}\eta^{(j)}$, we have

\vspace{-5mm}
\begin{align}
    &\hspace{-4mm}\sum_{t=1}^{T} \left(r_{a^\star,\,t} - r_{t}^{\top}\distfsf_t^{(j)} \right)\nonumber\\\label{aux1:8}
    \leq& \frac{\Bar{P}(T)\log{|\calV|}}{\eta} + \frac{P(T)\log{T}+1}{\eta} + \frac{\Bar{P}(T)\log{|\calV|}+P(T)\log{\left(\frac{eT}{P(T)}\right)}+\log{\left(\frac{1}{\delta}\right)}}{2\gamma}+ \left(\frac{\eta}{2}+\gamma\right)|\calV|T + \left(\frac{\eta}{2}+\gamma\right)\frac{\log{\left(\frac{1}{\delta}\right)}}{2\gamma} \\
    = &\frac{2\Bar{P}(T)\log{|\calV|} + P(T)\log{T}+1 + P(T)\log{\left(\frac{eT}{P(T)}\right)}+\log{\left(\frac{1}{\delta}\right)}}{\eta} + \eta |\calV|T + \log{\left(\frac{1}{\delta}\right)} \\
    = &\frac{2\Bar{P}(T)\log{|\calV|} + 2P(T)\log{T}+ 1 + P(T) - P(T)\log{P(T)}}{\eta} + \eta |\calV|T + \left(\frac{1}{\eta}+1\right)\log{\left(\frac{1}{\delta}\right)} \\\label{aux1:9}
    \leq &\frac{2\Bar{P}(T)\log{|\calV|} + 2P(T)\log{T} + 2\log{T}}{\eta} + \eta |\calV|T + \left(\frac{1}{\eta}+1\right)\log{\left(\frac{1}{\delta}\right)} \\
    = &\frac{2\Bar{P}(T)\log{(|\calV|T)}}{\eta} + \eta |\calV|T + \left(\frac{1}{\eta}+1\right)\log{\left(\frac{1}{\delta}\right)} 
\end{align}
holds with probability at least $1-\delta$, where \cref{aux1:8} holds from eq. (17) of \cite[Appendix B.1]{matsuoka2021tracking}, and \cref{aux1:9} holds because $1+P(T)-P(T)\log{P(T)}<2\log{T}$ for $T\geq 3$. 
By the definition of $\{\eta^{(j)}\}_{j\myin[J]}$, there always exists a $j\in[J]$ such that 
\begin{equation}
    \frac{\eta^{(j)}}{2} \leq \sqrt{\frac{\Bar{P}(T)\log{(|\calV|T)}}{|\calV|T}} \leq \eta^{(j)}.
\end{equation}For this $j$, since $|\calV|\geq 2$ and $T\geq 3$, we know $1<\log{(|\calV|T)}$, and, thus,  with probability at least $1-\delta$,

\vspace{-4mm}
\begin{align}
    &\hspace{-4mm}\sum_{t=1}^{T} \left(r_{a^\star,\,t} - r_{t}^{\top}\distfsf_t^{(j)}\right) \\
    \leq &\;\frac{2\Bar{P}(T)\log{(|\calV|T)}}{\eta} + \eta |\calV|T + \left(\frac{1}{\eta}+1\right)\log{\left(\frac{1}{\delta}\right)} \\\nonumber
    \leq & \;2\Bar{P}(T)\log{(|\calV|T)}\sqrt{\frac{|\calV|T}{\Bar{P}(T)\log{(|\calV|T)}}} + 2\sqrt{\Bar{P}(T)|\calV|T\log{(|\calV|T)}} +\left(\sqrt{\frac{|\calV|T}{\Bar{P}(T)\log{(|\calV|T)}}} + 1\right)\log{\left(\frac{1}{\delta}\right)} \\
    \label{aux1:10}
    \leq & \;4\sqrt{\Bar{P}(T)|\calV|T\log{(|\calV|T)}} + \left(\sqrt{\frac{|\calV|T}{\Bar{P}(T)\log{(|\calV|T)}}} + 1\right)\log{\left(\frac{1}{\delta}\right)}.
\end{align} 

Hence, \cref{aux1:10} holds true for $T\geq3$ and $|\calV|\;\geq 2$.

In all, combining \cref{aux1:10,aux1:MWU_bound}, the following steps hold true for $T\geq3$ and $|\calV|\;\geq 2$ with probability at least $1-\delta$:
\begin{align}
    &\sum_{t=1}^{T}\left( r_{a^\star,\,t} - r_{t}^{\top}p_t\right) \nonumber\\
    &= \sum_{t=1}^{T} \left(r_{a^\star,\,t} - r_{t}^{\top}\distfsf_t^{(j)} \right)+\sum_{t=1}^{T}\left( r_{t}^{\top}\distfsf_t^{(j)} - r_{t}^{\top}p_t\right) \nonumber\\\label{aux1:11}
    &\leq 4\sqrt{\Bar{P}(T)|\calV|T\log{(|\calV|T)}} + 2\sqrt{2T\log{J}} + \left(\sqrt{\frac{|\calV|T}{\Bar{P}(T)\log{(|\calV|T)}}} + 1\right)\log{\left(\frac{1}{\delta}\right)} \\
    &\leq 4\sqrt{\Bar{P}(T)|\calV|T\log{(|\calV|T)}} + 4\sqrt{T\log{(1+\log{T})}} + \left(\sqrt{\frac{|\calV|T}{\Bar{P}(T)\log{(|\calV|T)}}} + 1\right)\log{\left(\frac{1}{\delta}\right)} \\
    &\leq 4\sqrt{2T\Big(\Bar{P}(T)|\calV|\log{(|\calV|T)} + \log{(1+\log{T})\Big)}} + \left(\sqrt{\frac{|\calV|T}{\Bar{P}(T)\log{(|\calV|T)}}} + 1\right)\log{\left(\frac{1}{\delta}\right)},
\end{align} where \cref{aux1:11} holds because $\log{J}\leq2\log{(1+\log{T})}$ for $T\geq 3$. \qed

%% file: App-2.tex
\vspace{1cm}
\section{Proof of \Cref{th:half}}\label{app:BSG}
We denote by $\solopt_{i-1,\,t}$ the optimal solution set for the first $i-1$ agents at time step $t$. Then, we have:{
\begin{align}
    &\hspace{-2.1cm}\sum_{t=1}^{T} f_t(\,\solopt_t\,) \nonumber\\
    &\hspace{-2.1cm}\leq \sum_{t=1}^{T} f_t(\,\solopt_t\cup\solbsg_t\,)\label{aux22:1} \\\label{aux22:2}
    &\hspace{-2.1cm}= \sum_{t=1}^{T} f_t(\,\solbsg_t\,) + \sum_{t=1}^{T} \sum_{i\myin\calN} f_t(a_{i,\,t}^\opt\,|\,\solopt_{i-1,\,t}\cup\solbsg_t\,)\\\label{aux22:3}
    &\hspace{-2.1cm}\leq \sum_{t=1}^{T} f_t(\,\solbsg_t\,) + \sum_{t=1}^{T} \sum_{i\myin\calN} f_t(a_{i,\,t}^\opt\,|\,\solbsg_{i-1,\,t}\,)
\end{align}
\begin{align}
    \label{aux22:4}&= 2\sum_{t=1}^{T} f_t(\,\solbsg_t\,)+\sum_{t=1}^{T} \sum_{i\myin\calN} f_t(a_{i,\,t}^\opt\,|\,\solbsg_{i-1,\,t}\,) - f_t(a_{i,\,t}^\banalg\,|\,\solbsg_{i-1,\,t}\,)\\
    \label{aux22:5}
    &= 2\sum_{t=1}^{T} f_t(\,\solbsg_t\,) + \sum_{t=1}^{T} \sum_{i\myin\calN} r_{a^\opt,\,t}^{\hspace{.5pt}(i)} - r_{a^\banalg,\,t}^{\hspace{.5pt}(i)},
\end{align}}where \cref{aux22:1} holds from the monotonicity of $f_t$; \cref{aux22:2,aux22:4} are proved by telescoping the sums; \cref{aux22:3} holds from the submodularity of $f_t$; and \cref{aux22:5} holds from the definition of $r_{a^\banalg,\,t}^{\hspace{.5pt}(i)}$ (\Cref{alg:online}'s line 14). Now:
{
\begin{align}
    &\mathbb{E}\Bigl[\scenario{Tracking}\text{-}\scenario{Regret}_T^{(1/2)}(\,\solbsg\,)\Bigr] \nonumber    \\\label{aux2:1}
    &= \mathbb{E}\Bigl[\frac{1}{2}\sum_{t=1}^{T} f_t(\,\solopt_t\,) - \sum_{t=1}^{T}f_t(\,\solbsg_t\,)\Bigr]\\\label{aux2:2}
    &\leq \frac{1}{2}\sum_{t=1}^{T}  \sum_{i\myin\calN} \mathbb{E} \left(r_{a^\opt,\,t}^{\hspace{.5pt}(i)} - r_{a^\banalg,\,t}^{\hspace{.5pt}(i)}\right)\\\label{aux2:3}
    &= \frac{1}{2}\sum_{t=1}^{T}  \sum_{i\myin\calN} \left(r_{a^\opt,\,t}^{\hspace{.5pt}(i)} - r_{t}^{\hspace{.5pt}(i)\top}\distfsf_t^{(i)}\right)\\\label{aux2:4} 
    &\leq \frac{1}{2}\Bigg[4 \sum_{i\myin\calN} \sqrt{2T\Big(\Bar{\Delta}_i(T)|\calV_i|\log{(|\calV_i|T)}+\log{(1+\log{T})}\Big)}+ \sum_{i\myin\calN}\left(\sqrt{\frac{|\calV_i|T}{\Bar{\Delta}_i(T)\log{(|\calV_i|T)}}} + 1\right)\log{\left(\frac{1}{\delta}\right)}\Bigg]\\\label{aux2:5}
    &\leq 2 \sqrt{2|\calN|T\sum_{i\myin\calN}\Big(\Bar{\Delta}_i(T)|\calV_i|\log{(|\calV_i|T)}+\log{(1+\log{T})}\Big)} + \frac{1}{2}\left(\sqrt{|\calN|T\sum_{i\myin\calN}\frac{|\calV_i|}{\Bar{\Delta}_i(T)\log{(|\calV_i|T)}}} + |\calN|\right)\log{\left(\frac{1}{\delta}\right)}\\\label{aux2:6}
    &\leq 2 \sqrt{2|\calN|T\Big((\Delta(T)+ |\calN|){|\bar{\calV}|\log{({|\bar{\calV}|}T)}}+|\calN|\log{(1+\log{T})}\Big)} + \frac{1}{2}\left(\sqrt{|\calN|T\sum_{i\myin\calN}\frac{|\calV_i|}{\Bar{\Delta}_i(T)\log{(|\calV_i|T)}}} + |\calN|\right)\log{\left(\frac{1}{\delta}\right)},
\end{align}}with probability at least $1-\delta$, where \cref{aux2:1} holds from \cref{eq:regret_simplified}; 
\cref{aux2:2} holds from \cref{aux22:5}; 
\cref{aux2:3} holds from the internal randomness of \expsix{i}; \cref{aux2:4} holds from~\cite[Corollary 1]{matsuoka2021tracking}; \cref{aux2:5} holds from the Cauchy–Schwartz inequality; {and  
\cref{aux2:6} holds} true since $\sum_{i\myin\calN} \Delta_i(T) =\Delta(T)$. 
\qed

%% file: App-3.tex
\vspace{1cm}
\section{{Proof of Monotonicity and Submodularity of Function~\eqref{eq:distance}}}
Because the addition of multiple non-decreasing submodular functions results to non-decreasing submodular functions, it suffices to prove that the function $f(\calS)=-1\,/\,\left(\,\sum_{s\in\calS} \;{1}/{s}\,\right)$, {$\calS\in 2^{\mathbb{R}_+}$} is non-decreasing and submodular, where $f(\emptyset)={-\infty}$. 
We start by proving $f$'s monotonicity: consider {$\calA\subseteq \calB\in 2^{\mathbb{R}_+}$}, then we have  $-\sum_{a\in\calA} {1}/{a} \geq -\sum_{b\in\calB}  {1}/{b}$, and thus $f(\calA) \leq f(\calB)$. We now prove $f$'s submodularity:  consider finite and disjoint $\calB_1\in 2^{\mathbb{R}_+}$ and $\calB_2\in 2^{\mathbb{R}_+}$, and an arbitrary non-zero real number $s$. Set $B_1\triangleq\sum_{b_1\in\calB_1}{1}/{b_1}$ and   $B_2\triangleq\sum_{b_2\in\calB_2}{1}/{b_2}$; then, 
$$\frac{1}{\calB_1+\calB_2} - \frac{1}{\calB_1+\calB_2+1/s} \leq \frac{1}{\calB_1} - \frac{1}{\calB_1+1/s},$$where the equality is taken when $\calB_2=\emptyset$. Therefore, $f(\,\calB_1\cup\calB_2\cup\{s\}) - f(\,\calB_1\cup\calB_2) \leq f(\,\calB_1\cup\{s\}) - f(\,\calB_1)$, which proves $f$'s submodularity. \qed